\documentclass[]{interact}

\usepackage{epstopdf}
\usepackage{bm}
\usepackage{blkarray}
\usepackage{url}
\usepackage{amsmath}
\usepackage{comment}
\usepackage{dcolumn}
\newcolumntype{d}[1]{D{.}{.}{#1} }
\usepackage{booktabs}
\usepackage{multirow}
\usepackage{titlesec}

\DeclareMathOperator{\sinc}{sinc}
\usepackage[numbers,sort&compress]{natbib}
\bibpunct[, ]{[}{]}{,}{n}{,}{,}

\makeatletter
\def\NAT@def@citea{\def\@citea{\NAT@separator}}
\makeatother

\usepackage{tikz,xcolor,hyperref}

\definecolor{linkcolour}{HTML}{000066}	
\hypersetup{colorlinks,breaklinks,
	urlcolor=linkcolour, 
	linkcolor=linkcolour,
	citecolor=linkcolour}

\definecolor{lime}{HTML}{A6CE39}
\DeclareRobustCommand{\orcidicon}{
	\begin{tikzpicture}
		\draw[lime, fill=lime] (0,0) 
		circle [radius=0.16] 
		node[white] {{\fontfamily{qag}\selectfont \tiny ID}};
		\draw[white, fill=white] (-0.0625,0.095) 
		circle [radius=0.007];
	\end{tikzpicture}
	\hspace{-2mm}
}

\newcommand{\orcidauthorA}{\href{https://orcid.org/0000-0002-1554-3820}{\orcidicon}  } 
\newcommand{\orcidauthorB}{\href{https://orcid.org/0000-0001-5864-9636}{\orcidicon}} 

\newcommand{\subsubsubsection}[1]{\paragraph{#1}\mbox{}\\}
\begin{document}

\title{Random matrix theory of polarized light scattering in disordered media}

\author{
\name{Niall Byrnes{\orcidauthorA} and Matthew R. Foreman{\orcidauthorB}\textsuperscript{1}\thanks{\textsuperscript{1}Corresponding author:  matthew.foreman@imperial.ac.uk}}
\affil{Blackett Laboratory, Department of Physics, Imperial College London, Prince Consort Road, London SW7 2AZ, United Kingdom}
}

\maketitle

\begin{abstract}
In this work we present a method for generating random matrices describing electromagnetic scattering from disordered media containing dielectric particles with prescribed single particle scattering characteristics. Resulting scattering matrices automatically satisfy the physical constraints of unitarity, reciprocity and time reversal, whilst also incorporating the polarization properties of electromagnetic waves and scattering anisotropy. Our technique therefore enables statistical study of a variety of polarization phenomena, including depolarization rates and polarization-dependent scattering by chiral particles. In this vein, we perform numerical simulations for media containing isotropic and chiral spherical particles of different sizes for thicknesses ranging from the single to multiple scattering regime and discuss our results, drawing comparisons to established theory.   
\end{abstract}

\section{Introduction}
Complex, disordered media are ubiquitous in nature, from cosmic dust in the interstellar medium to tissue in the brain \cite{Escobar_Cerezo_2017, Favre-Bulle2015}. When light interacts with such media, multiple scattering can cause severe deterioration of the spatio-temporal structure of the incident field through randomization of amplitude, phase and polarization state.  Multiple scattering therefore can heavily degrade optical information \cite{goodman2007speckle,Byrnes_2020} posing significant challenges in many scientific disciplines, including telecommunications, remote sensing, astronomy, medical diagnostics and optical imaging \cite{thevenaz2011advanced, rees2013physical,tyson2012principles,tuchin2007tissue, 10.1117/1.3652896}.  
A detailed understanding of the transport of light in complex systems is paramount to overcoming limitations imposed by multiple scattering, therefore necessitating development of accurate modelling tools.

Theoretically modeling multiple scattering of polarized light is notoriously difficult. While in principle the scattered field follows exactly from Maxwell's equations, numerous approximations are generally required to render the mathematics tractable \cite{mishchenko2006multiple}. Numerical solutions of Maxwell's equations have been performed for scattering systems of limited size, typically with dimensions on the order of ten wavelengths, using the T-matrix method, time domain simulations and the coupled dipole approximation \cite{Mishchenko:07, https://doi.org/10.1029/2005RS003408, PhysRevE.77.066709}. A popular alternative approach for modeling low-density scattering media is the radiative transfer equation (and its vectorial counterpart), which predicts the specific intensity (Stokes vector) at a far field measurement point \cite{10.1117/1.JBO.21.7.071114}. The radiative transfer equations are frequently solved numerically using Monte Carlo approaches that trace rays, thought of as `photons', through the scattering medium \cite{PhysRevLett.95.213901, Bartel:00, Antonelli:10}. The path of each photon is simulated as a random walk where scattering events occur at random positions at which the photon wavevector is updated probabilistically using a specified phase function. The polarization state of a photon after scattering can then be updated using an amplitude scattering matrix that can be customized according to the type of particle being modeled. One drawback of the Monte Carlo technique is speed; while much faster than directly solving Maxwell's equations, a large amount of computation is required to estimate statistical quantities with high accuracy. Simulations must also be repeated when photons are injected with different angles of incidence. In addition, traditional Monte Carlo methods are unable to reproduce correlations such as the memory effect, although more recent studies have begun to address this problem \cite{Shen:17, 10.1145/3306346.3322950}. 

Scattering matrices, and the closely related transmission, reflection and transfer matrices, provide an alternative description of a scattering medium \cite{RevModPhys.89.015005}. Practically, the scattering matrix (or other related matrices) can be determined through sequential measurements using a spatial light modulator to control the different degrees of freedom of an electric field \cite{Miller:19, doi:10.1063/1.5048493, app7060568, doi:10.1063/1.4984209, Tripathi:12}. Once known, the scattering matrix determines the response of a medium to an arbitrary incident field and enables the design of incident wavefronts that, rather than being distorted by multiple scattering, are tightly focused or strongly transmitted well beyond the ballistic regime \cite{Xu:17, PhysRevLett.104.100601, Mosk2012}. In addition, when viewed statistically, correlations between different matrix elements can embody phenomena such as the optical memory effect \cite{Bertolotti2012, Katz2014, Judkewitz2015,Edrei2016} or transmission-reflection correlations, which have been exploited for imaging when the transmitted field is inaccessible \cite{PhysRevA.92.033827, PhysRevX.8.021041, Paniagua-Diaz:19}.

For complex media, the scattering matrix can be treated as a random matrix sampled from some suitable matrix ensemble. This matrix ensemble is the set of all scattering matrices corresponding to all possible microscopic configurations of a system with a given set of macroscopic properties, such as particle density, mean particle size etc. It is well known that for any non-absorbing system, physically admissible scattering matrices are constrained to be unitary due to energy conservation, with an additional symmetry constraint imposed when reciprocity or time reversal symmetry holds \cite{forrester2010log, PhysRevResearch.3.013129}. The earliest random matrix models for the scattering matrix, namely the circular ensembles, are based on the use of a uniform (Haar measure) distribution over the unitary group \cite{akemann2018oxford}. A more sophisticated random matrix model is captured in the DMPK equation, which describes the statistical evolution of the singular values of the transmission matrix for a random medium \cite{RevModPhys.69.731, mello2004quantum}. While sufficient for revealing universal properties of disordered media, such as the existence of highly transmitting open eigenchannels \cite{RevModPhys.69.731}, these models are largely limited to purely isotropic scattering media and contain no adjustable parameters for exploring the multitude of phenomena exhibited by real systems. Moreover, random matrix models typically only consider scalar waves and thus can not describe polarization dependent effects. Generalizations of the DMPK equation have been proposed, but are typically expressed in terms of correlations between the singular values and vectors of the transmission and reflections matrices \cite{PhysRevLett.82.4272, PhysRevB.66.115318, Douglas_2014}. These variables are physically unintuitive and their statistical properties can be difficult to relate to those of the elements of the scattering matrix. Monte Carlo transfer matrix simulations for disordered waveguides with non-isotropic scattering have also been performed, but to our knowledge have also not yet incorporated polarization effects, which are particularly important for optical scattering \cite{PhysRevE.75.031113}. In this work we address these limitations by presenting a method for numerically generating optical scattering matrices for random media of arbitrary thicknesses, incorporating the polarization properties of light. Our method requires the prescription of the single scattering properties of the particles that constitute the random medium and uses a matrix cascade approach to simulate the multiple scattering regime. We consider sparse distributions of randomly positioned particles such that each scatterer is in the far field of all other scatterers. Arbitrary fields are expressed using a discrete angular spectrum of plane waves, which facilitates the description of non-planar wavefronts and allows the theory to be expressed in terms of the scattering of plane waves, for which the literature is abundant.

The content of this paper is organized as follows. In Section \ref{theory}, we cover the background theory relevant to the model. We begin in Section \ref{scatmat} by defining the scattering matrix and deriving expressions for its elements in the single scattering regime.  In Section \ref{statistics}, we detail the statistical properties of the scattering matrix and derive expressions for the mean, covariance matrix and pseudo-covariance matrix associated with the scattering matrix elements. The issue of enforcing necessary matrix symmetries on randomly generated matrices is briefly discussed in Section \ref{symmetries}. We present numerical simulations of random media consisting of dielectric spheres in Section \ref{simulations}. Specifically, our method is explained in Section \ref{method}, with results presented in Section \ref{results}. In particular, we present statistical data for the transmission eigenvalues as well as the scattered intensity, DoP, retardance and diattenuation for different outgoing plane wave directions. For all of our results, we discuss their physical interpretations, drawing comparisons to established theory. We end with a summary and conclusion of our work.  

\section{Theory}\label{theory}
In this section we give a comprehensive description of the theoretical model used in our simulations. We begin by setting out the problem we wish to study and derive expressions for the scattering matrix elements in the single scattering regime. We then discuss the statistical properties of the scattering matrix elements, which can be related to the properties of the individual scatterers in the medium. Finally, we discuss how the matrix symmetries imposed by energy conservation and reciprocity are enforced.

\subsection{The scattering matrix}\label{scatmat}
Consider a slab of thickness $\Delta L$, bounded by the planes $z = -\Delta L/2$ and $z = \Delta L/2$ and infinite in transverse extent. Suppose that the slab contains $N$ dielectric particles distributed sparsely enough so that each particle is in the far field (defined rigorously below) of all the others. We assume that the boundaries of the slab are non-reflective so that scattering only occurs due to the presence of the particles within the slab. Suppose that the slab is illuminated by a right-propagating plane wave (`right' henceforth meaning in the positive $z$ direction) with wavevector $\mathbf{k}_i = (k_{ix},k_{iy},k_{iz})^\mathrm{T}$ ($\mathrm{T}$ denoting the transpose operator) where $k_{iz} > 0$, $|\mathbf{k}_i| = k = 2\pi/\lambda$ and $\lambda$ is the wavelength. The complex representation of the electric field associated with the incident wave at position $\mathbf{r}$ is given by
\begin{equation}
	\mathbf{E}_i(\mathbf{r}) = \mathbf{E}_0e^{i\mathbf{k}_i \cdot \mathbf{r}} = \int\delta(\bm\kappa - \bm\kappa_i)\mathbf{E}_0 e^{i(\bm{\kappa}\cdot \bm{\rho} + k_z z)}\mathrm{d}k_x \mathrm{d}k_y,
\end{equation}
where $\bm{\kappa} = (k_x, k_y)^\mathrm{T}$ and $\bm{\rho} = (x,y)^\mathrm{T}$ are the transverse wavevector and transverse position vector, $k_z = (k^2 - k_x^2 - k_y^2)^{1/2}$ and $\delta$ is the Dirac delta function. The vector $\mathbf{E}_0$ is constant and characterizes the polarization state of the incident wave.

Suppose now that the slab thickness $\Delta L$ is sufficiently small so that the total scattered field can be assumed to be composed of only single scattering contributions from each particle. If the centre of the $p$'th particle is located at position $\mathbf{r}_p$, its single scattering contribution to the total field $\mathbf{E}_p$ in the far field (i.e. $k|\mathbf{r}-\mathbf{r}_p| \gg 1$) is given by 
\begin{align}
\mathbf{E}_p(\mathbf{r}) = \frac{e^{ik|\mathbf{r}-\mathbf{r}_p|}}{ik|\mathbf{r}-\mathbf{r}_p|}\mathbf{A}_p(\Delta\mathbf{r}, \mathbf{u}_i)\mathbf{E}_0 e^{i\mathbf{k}_i\cdot\mathbf{r}_p},\label{farfieldsphere}
\end{align}
where $\Delta\mathbf{r} = (\mathbf{r}-\mathbf{r}_p)/|\mathbf{r}-\mathbf{r}_p|$ and $\mathbf{u}_i = \mathbf{k}_i/k$ are unit vectors \cite{bohren2008absorption}. The $3\times 3$ matrix $\mathbf{A}_p(\Delta\mathbf{r},\mathbf{u}_i)$, which depends on the shape, size, orientation and morphology of the scatterer, describes the transformation of the polarization state of the incident field to that of the scattered field in the far field observation direction $\Delta\mathbf{r}$. Eq. (\ref{farfieldsphere}) admits an angular spectrum representation, which is given by
\begin{equation}
\mathbf{E}_{p}(\mathbf{r}) = \int \frac{\mathbf{A}_p(\mathbf{u},\mathbf{u}_i)e^{i(\mathbf{k}_i - \mathbf{k})\cdot\mathbf{r}_p}}{2\pi k k_z}\mathbf{E}_0 e^{i(\bm{\kappa}\cdot \bm{\rho} + k_z z)}\mathrm{d}k_x\mathrm{d}k_y,\label{Escatangspec}
\end{equation}
where now $\mathbf{r} = (\bm\rho, z)^\mathrm{T}$, $\mathbf{k} = (\bm\kappa, k_z)^\mathrm{T}$ and $\mathbf{u} = \mathbf{k}/k$ \cite{mandel1995optical}. Since $\mathbf{r}$ lies in far field of the scatterer, the domain of integration in Eq. (\ref{Escatangspec}) is restricted to the set of all wavevectors for which $|\bm{\kappa}| < k$, i.e. homogeneous plane waves. Considering now the total electric field on the planar boundaries of the scattering medium, we find the expressions
\begin{align}
	\mathbf{E}(\bm\rho, \Delta L/2) &=  \int \Big[\delta(\bm{\kappa}- \bm{\kappa}_i)\mathbb{I}_3 + \sum_{p=1}^N\frac{\mathbf{A}^t_p(\bm{\kappa},\bm{\kappa}_i)}{2\pi kk_z}e^{i(\mathbf{k}_i - \mathbf{k})\cdot\mathbf{r}_p}\Big]\mathbf{E}_0 e^{i(\bm{\kappa}\cdot \bm{\rho} + k_z \Delta L/2)}\mathrm{d}k_x\mathrm{d}k_y\label{angspec1},\\
\begin{split}
	\mathbf{E}(\bm\rho, -\Delta L/2) &= \int \sum_{p=1}^N\frac{\mathbf{A}^r_p(\bm{\kappa},\bm{\kappa}_i)}{2\pi k k_z}e^{i(\mathbf{k}_i -\widetilde{\mathbf{k}})\cdot\mathbf{r}_p}\mathbf{E}_0 e^{i(\bm{\kappa}\cdot \bm{\rho} - k_z (-\Delta L/2))}\mathrm{d}k_x\mathrm{d}k_y\\
	&\quad\quad\quad\quad\quad\quad\quad\quad\quad\quad\quad\quad\quad\quad\quad\quad\quad\quad+\, \mathbf{E}_0e^{i(\bm{\kappa}_i\cdot\bm\rho + k_{iz}(-\Delta L/2))},\label{angspec2}
\end{split}
\end{align} 
where $\mathbb{I}_n$ is the $n \times n$ identity matrix and we have defined $	\mathbf{A}^t_p(\bm{\kappa},\bm{\kappa}_i) = \mathbf{A}_p(\mathbf{u},\mathbf{u}_i)$ and $  \mathbf{A}^r_p(\bm{\kappa},\bm{\kappa}_i) = \mathbf{A}_p(\widetilde{\mathbf{u}},\mathbf{u}_i)$. We use a tilde to denote a wavevector with negative $z$ component, i.e. if $\mathbf{u}= (u_x,u_y,u_z)^\mathrm{T}$ with $u_z = (|\mathbf{u}|^2 - u_x^2-u_y^2)^{1/2} > 0$, then $\widetilde{\mathbf{u}}= (u_x,u_y,-u_z)^\mathrm{T}$. Assuming that the planar boundaries also lie in the far field of every particle within the scattering medium, evanescent wave contributions to the integrals in Eqs. (\ref{angspec1}) and (\ref{angspec2}) can also be neglected. 

In Eqs. (\ref{angspec1}) and (\ref{angspec2}), the matrices $\mathbf{A}^t_p$ and $\mathbf{A}^r_p$ are continuous functions of transverse wavevector. In reality, however, it is only possible to simulate the scattered field up to some minimal resolution. We hence construct discrete counterparts to Eqs. (\ref{angspec1}) and (\ref{angspec2}) by replacing the integrals with sums over a finite set of wavevectors. We define the set $K = \{-\bm{\kappa}_{N_k},\hdots,-\bm{\kappa}_2,-\bm{\kappa}_1,\mathbf{0},\bm{\kappa}_1,\bm{\kappa}_2,\hdots,\bm{\kappa}_{N_k}\},\label{modes}$ which consists of $N_k$ transverse wavevectors (henceforth referred to as `modes') together with their additive inverses and the two component zero vector $\mathbf{0} = (0, 0)^\mathrm{T}$, which corresponds to the wavevector $(0,0,k)^\mathrm{T}$. For each mode $\bm\kappa_i \in K$, we also define an associated weight $w_i$, where $\sum_i w_i = \pi k^2$, so that for any function $f$ we have the cubature scheme
\begin{equation}
\int f(\bm\kappa)\mathrm{d}k_x \mathrm{d}k_y \approx \sum_i f(\bm\kappa_i) w_i.\label{cubature}
\end{equation}
Naturally, increasing the number of modes improves the accuracy of Eq. (\ref{cubature}), albeit at the expense of an increase in computation. Many different choices of modes and weights are possible in principle, and the optimal choice of cubature scheme may depend non-trivially on the forms of $\mathbf{A}^t_p$ and $\mathbf{A}^r_p$. In this work, we used modes distributed on a Cartesian grid in $k$ space, each having an equal weight given by $w_i = w = \pi k^2/(2N_k+1)$ for all $i$. Finally, we note that it is necessary to choose modes in inverse pairs to fully exploit scattering reciprocity \cite{PhysRevResearch.3.013129}.

Given a cubature scheme defined as above, Eqs. (\ref{angspec1}) and (\ref{angspec2}) can be discretized to
\begin{align}
\mathbf{E}(\bm\rho, \Delta L/2) &= \sum_{j = -N_k}^{N_k}\mathbf{t}(\bm{\kappa}_j, \bm{\kappa}_i)\mathbf{E}_0 e^{i(\bm{\kappa}_j\cdot \bm{\rho} + k_{jz} \Delta L/2)}w\label{discrete1},\\
\mathbf{E}(\bm\rho, -\Delta L/2) &= \mathbf{E}_0e^{i(\bm{\kappa}_i\cdot\bm\rho - k_{iz}\Delta L/2)} + \sum_{j = -N_k}^{N_k}\mathbf{r}(\bm\kappa_j, \bm\kappa_i)\mathbf{E}_0 e^{i(\bm{\kappa}_j\cdot \bm{\rho} + k_
	{jz} \Delta L/2)}w,\label{discrete2}\
\end{align} 
where
\begin{align}
\mathbf{t}(\bm{\kappa}_j, \bm{\kappa}_i) &= \frac{\delta_{ij}}{w}\mathbb{I}_3 + \frac{1}{2\pi kk_{jz}}\sum_{p=1}^N\mathbf{A}^t_p(\bm{\kappa}_j,\bm{\kappa}_i)e^{i(\mathbf{k}_i - \mathbf{k}_j)\cdot\mathbf{r}_p},\\
\mathbf{r}(\bm{\kappa}_j, \bm{\kappa}_i) &= \frac{1}{2\pi kk_{jz}}\sum_{p=1}^N\mathbf{A}^r_p(\bm{\kappa}_j,\bm{\kappa}_i)e^{i(\mathbf{k}_i - \widetilde{\mathbf{k}}_j)\cdot\mathbf{r}_p}\
\end{align}	
are transmission and reflection matrices. Note that we have replaced the differential product $\mathrm{d}k_x\mathrm{d}k_y$ with $w$ and the delta function $\delta(\bm\kappa - \bm\kappa_i)$ with the normalized Kronecker delta $\delta_{ij}/w$. For the transverse wavevectors, we use integer subscripts where negative values correspond to modes listed in $K$ with a negative sign, e.g. $\bm\kappa_{-1} = -\bm\kappa_1$, and $\bm\kappa_0 = \mathbf{0}$. 

As there are no sources in the planes $z=-\Delta L/2$ and $z=\Delta L/2$, it follows from the Maxwell equation $\nabla\cdot\mathbf{E} = 0$ that only four of the nine elements of the transmission and reflection matrices are independent \cite{mishchenko2006multiple}. These matrices may therefore be reduced to $2\times 2$ matrices, which is facilitated by introducing the standard spherical polar coordinates basis vectors
\begin{equation}
	\mathbf{e}_k(\bm{\kappa}, k_z) = \frac{\mathbf{k}}{k},\quad	\quad
	\mathbf{e}_\phi(\bm{\kappa}, k_z) = \frac{\hat{\mathbf{z}} \times  \mathbf{e}_k}{|\hat{\mathbf{z}} \times  \mathbf{e}_k|},\quad\ \quad
	\mathbf{e}_\theta(\bm{\kappa}, k_z) =  \frac{\mathbf{e}_\phi \times \mathbf{e}_k}{|\mathbf{e}_\phi \times \mathbf{e}_k|}.\label{polar}
\end{equation}
For the special cases $\mathbf{e}_k = \pm\hat{\mathbf{z}}$, we set $\mathbf{e}_\phi = \pm \hat{\mathbf{y}}$. We define the reduced $2\times 2$ transmission and reflection matrices to be $\mathbf{t}_{(j,i)}$ and $\mathbf{r}_{(j,i)}$ whose elements are defined by
\begin{align}
	t_{(j,i)mn} &= \mathbf{e}_m^{\mathrm{T}}(\bm\kappa_j, k_{jz})\mathbf{t}(\bm\kappa_j, \bm\kappa_i)\mathbf{e}_n(\bm\kappa_i, k_{iz}),\\
	r_{(j,i)mn} &= \mathbf{e}_m^{\mathrm{T}}(\bm\kappa_j, -k_{jz})\mathbf{r}(\bm\kappa_j, \bm\kappa_i)\mathbf{e}_n(\bm\kappa_i, k_{iz}),
\end{align}
where $m$ and $n$ stand for either $\theta$ or $\phi$. Finally, for mathematical convenience (see Ref. \cite{PhysRevResearch.3.013129} for more details), we normalize the transmission and reflection matrices to $\bar{\mathbf{t}}_{(j,i)}$ and $\bar{\mathbf{r}}_{(j,i)}$, which are given by
\begin{align}
	\bar{\mathbf{t}}_{(j,i)} &=  \sqrt{\frac{k_{jz}}{k_{iz}}}\mathbf{t}_{(j,i)}w = \delta_{ij}\mathbb{I}_2 + C_{ji}\sum_{p=1}^N\mathbf{A}^t_{p(j,i)}e^{i(\mathbf{k}_i - \mathbf{k}_j)\cdot\mathbf{r}_p}\label{t},\\
	\bar{\mathbf{r}}_{(j,i)} &= \sqrt{\frac{k_{jz}}{k_{iz}}}\mathbf{r}_{(j,i)}w =  C_{ji}\sum_{p=1}^N\mathbf{A}^r_{p(j,i)}e^{i(\mathbf{k}_i - \widetilde{\mathbf{k}}_j)\cdot\mathbf{r}_p},\label{r}
\end{align}
where $C_{ji} = w /(2\pi k \sqrt{k_{jz}k_{iz}})	$ and $\mathbf{A}^t_{p(j,i)}$ and $\mathbf{A}^r_{p(j,i)}$ are defined analogously to $\mathbf{t}_{(j,i)}$ and $\mathbf{r}_{(j,i)}$. 

The indices $i$ and $j$, which label the matrices $\bar{\mathbf{t}}_{(j,i)}$ and $\bar{\mathbf{r}}_{(j,i)}$, span from $-N_k$ to $N_k$, meaning there are a total of $(2N_k+1)^2$ transmission and reflection matrices. We may form an overall transmission and reflection matrix by concatenating $2\times 2$ blocks $\bar{\mathbf{t}}_{(j,i)}$ and $\bar{\mathbf{r}}_{(j,i)}$ for all pairs of incoming and outgoing modes taken from the set $K$. Specifically, we define $\bar{\mathbf{t}}$ (and $\bar{\mathbf{r}}$ analogously) to be the block matrix
\begin{equation}
	\bar{\mathbf{t}} = \;\begin{blockarray}{ccccccc}
		\begin{block}{(ccccccc)}
			\bar{\mathbf{t}}_{(-N_k,-N_k)} & \cdots & \bar{\mathbf{t}}_{(-N_k,-1)} & \bar{\mathbf{t}}_{(-N_k,0)} & \bar{\mathbf{t}}_{(-N_k,1)} & \cdots & \bar{\mathbf{t}}_{(-N_k,N_k)} \\
			\vdots & \ddots & \vdots & \vdots & \vdots & \ddots & \vdots \\
			\bar{\mathbf{t}}_{(-1,-N_k)} & \cdots & \bar{\mathbf{t}}_{(-1,-1)} & \bar{\mathbf{t}}_{(-1,0)} & \bar{\mathbf{t}}_{(-1,1)} & \cdots & \bar{\mathbf{t}}_{(-1,N_k)} \\
			\bar{\mathbf{t}}_{(0,-N_k)} & \cdots & \bar{\mathbf{t}}_{(0,-1)} & \bar{\mathbf{t}}_{(0,0)} & \bar{\mathbf{t}}_{(0,1)} & \cdots & \bar{\mathbf{t}}_{(0,N_k)} \\
			\bar{\mathbf{t}}_{(1,-N_k)} & \cdots & \bar{\mathbf{t}}_{(1,-1)} & \bar{\mathbf{t}}_{(1,0)} & \bar{\mathbf{t}}_{(1,1)} & \cdots & \bar{\mathbf{t}}_{(1,N_k)}  \\
			\vdots & \ddots & \vdots & \vdots & \vdots & \ddots & \vdots \\
			\bar{\mathbf{t}}_{(N_k,-N_k)} & \cdots & \bar{\mathbf{t}}_{(N_k,-1)} & \bar{\mathbf{t}}_{(N_k,0)} & \bar{\mathbf{t}}_{(N_k,1)} & \cdots & \bar{\mathbf{t}}_{(N_k,N_k)}  \label{bigt}\\
		\end{block}
	\end{blockarray}.
\end{equation}
The block $\bar{\mathbf{t}}_{(3,-2)}$, for example, describes transmission through the medium from mode $-\bm\kappa_{2}$ to mode $\bm\kappa_3$, i.e. from the incident right-propagating plane wave with wavevector $\mathbf{k}_{-2} = (-k_{2x}, -k_{2y}, k_{2z})^\mathrm{T}$ to that with wavevector $\mathbf{k}_3 = (k_{3x}, k_{3y}, k_{3z})^\mathrm{T}$. It is important to remember that in reflection, each outgoing plane wave component propagates to the left and has a wavevector with a negative $z$ component. The corresponding block of the reflection matrix $\bar{\mathbf{r}}_{(3,-2)}$ therefore describes the scattering from the same incident plane wave component to the left-propagating plane wave with wavevector $\widetilde{\mathbf{k}}_3 = (k_{3x}, k_{3y}, -k_{3z})^\mathrm{T}$.

Analogous expressions to those presented thus far can be derived for a left-propagating plane wave incident upon the right side of the scattering medium, yielding an additional pair of transmission and reflection matrices $\bar{\mathbf{t}}'$ and $\bar{\mathbf{r}}'$. Together, the matrices $\bar{\mathbf{t}}$, $\bar{\mathbf{r}}$, $\bar{\mathbf{t}}'$ and $\bar{\mathbf{r}}'$ form the normalized scattering matrix $\bar{\mathbf{S}}$, which is given by
\begin{equation}
	\bar{\mathbf{S}} = \begin{pmatrix}
		\bar{\mathbf{r}} &\bar{\mathbf{t}}' \\
		\bar{\mathbf{t}} & \bar{\mathbf{r}}'
	\end{pmatrix}.
\end{equation}
Put simply, the scattering matrix fully describes how waves incident upon the medium scatter into modes that propagate away from the medium, up to the resolution afforded by the mode discretization.

\subsection{Statistics of the scattering matrix elements}\label{statistics}
The expressions we have derived for the scattering matrix elements in Eqs. (\ref{t}) and (\ref{r}) are deterministic: if the locations and properties of all the scatterers are known, then in principle one can calculate the elements of $\bar{\mathbf{S}}$. In practice, however, the precise locations of every scatterer within the slab may be unknown and may vary considerably from one complex medium to another. It is therefore useful to think of $\bar{\mathbf{S}}$ as a random matrix. Observing Eqs. (\ref{t}) and (\ref{r}), we see that the `randomness' arises from two physical sources: the positions of the scatterers, which contribute to the complex exponential terms, and the morphological properties of the scatterers, i.e. shape, size, orientation etc., which contribute to the matrix factors $\mathbf{A}^t_p$ and $\mathbf{A}^r_p$. 

Observing Eqs. (\ref{t}) and (\ref{r}), with the exception of the diagonal elements of the transmission matrix ($i=j$), for which the argument of the complex exponential is always 0, the expressions for the transmission and reflection matrix elements  are essentially random phasor sums. Under rather general conditions, such expressions are known to be asymptotically Gaussian random variables as $N \to \infty$ \cite{goodman2015statistical}. For this to hold, we require the assumption that a scatterer's morphology is statistically independent of its position, which we shall take to be the case. We may therefore reasonably suppose that each of the matrix elements is marginally Gaussian distributed. It does not automatically follow that the the elements of $\bar{\mathbf{S}}$ follow a multivariate Gaussian distribution, but we shall nevertheless assume that this is the case. The statistics of a complex multivariate Gaussian distribution are fully described by three parameters: the mean, covariance matrix and pseudo-covariance matrix, expressions for which we shall now derive \cite{schreier2010statistical}.

Starting from Eq. (\ref{t}), we see that the mean value of $\bar{\mathbf{t}}_{(j,i)}$ is given by
\begin{equation}
\langle \bar{\mathbf{t}}_{(j,i)}\rangle = \delta_{ij}\mathbb{I}_2   +  NC_{ji}\langle\mathbf{A}^t_{(j,i)}\rangle\langle e^{i(\mathbf{k}_i - \mathbf{k}_j)\cdot\mathbf{r}}\rangle\label{tav},
\end{equation}
where we have used the independence of scatterer position and morphology. We have also assumed that each particle's $\mathbf{A}^t_p$ matrix is identically distributed, which allows us to drop the $p$ subscript. In order to compute the $\langle \exp[i(\mathbf{k}_i - \mathbf{k}_j)\cdot\mathbf{r}]\rangle$ term, it is first necessary to specify a probability distribution function for the particle position $\mathbf{r}$. We suppose that the particles are distributed uniformly in the slab so that the single particle distribution function is given by $p(\mathbf{r}) = 1/V$, where $V$ is the volume of the slab (momentarily taken to be finite). This assumption is reasonable given that each particle is in the far field of the others \cite{tsang2000scattering}.  Since the slab is infinite in transverse extent, both $N$ and $V$ are in fact infinite. We assume, however, that the particle density $n = N/V$ is finite and take the limit $N,V \to \infty$, holding $n$ constant. Therefore, we have
\begin{align}
\begin{split}
N\langle e^{i(\mathbf{k}_i - \mathbf{k}_j)\cdot\mathbf{r}}\rangle
&\to n\int_{-\Delta L/2}^{\Delta L/2}\int_{-\infty}^{\infty}\int_{-\infty}^{\infty}e^{i(\mathbf{k}_i - \mathbf{k}_j)\cdot\mathbf{r}}\mathrm{d}x\mathrm{d}y\mathrm{d}z\\
&= (2\pi)^2n\Delta L\sinc\Big((k_{iz} - k_{jz})\frac{\Delta L}{2}\Big)\delta(k_{ix} - k_{jx})\delta(k_{iy} - k_{jy}),\label{averaget} 
\end{split}
\end{align}
where $\sinc(x) = \sin(x)/x$. Replacing the delta functions in Eq. (\ref{averaget}) with normalized Kronecker delta symbols $[\delta(k_{ix} - k_{jx})\delta(k_{iy} - k_{jy}) \to \delta_{ij}/w]$, Eq. (\ref{tav}) ultimately becomes
\begin{equation}
\langle \bar{\mathbf{t}}_{(j,i)}\rangle  =\delta_{ij}\Bigg(\mathbb{I}_2 +  \frac{2\pi n \Delta L}{ k k_{iz}}\langle\mathbf{A}^t_{(j,i)}\rangle\Bigg).\label{tavfinal}
\end{equation}
It is evident from Eq. (\ref{tavfinal}) that the mean values of the transmission matrix elements are only non-zero for blocks lying on the diagonal of $\bar{\mathbf{t}}$, which describe forward scattering. The mean values of the reflection matrix elements can be calculated similarly. Starting from Eq. (\ref{r}), we arrive at the analogous result
\begin{equation}
\langle \bar{\mathbf{r}}_{(j,i)}\rangle = \delta_{ij}\frac{2\pi n \Delta L}{ kk_{iz}}\sinc\Big(k_{iz}\Delta L\Big)\langle\mathbf{A}^r_{(j,i)}\rangle\label{rav}.
\end{equation}
These values are also only non-zero for blocks lying on the diagonal of $\bar{\mathbf{r}}$. These blocks correspond to reflections of plane waves whose wavevectors transform according to $\mathbf{k} \to \widetilde{\mathbf{k}}$, i.e. scattering in the `specular reflection' direction. The sinc function in Eq. (\ref{rav}) is due to the randomness in $z$ position of the particles, which imparts a random phase onto each singly scattered component of the total field \cite{Barrera:03}.

Computing the covariances of the scattering matrix elements requires finding correlations of the form  $\langle \bar{t}_{(j,i)ba}\bar{t}^*_{(v,u)dc}\rangle$, where $i,j,u$ and $v$ refer to transverse wavevectors (taken from $K$) and $a,b,c$ and $d$ refer to polarization states ($\theta$ or $\phi$). Assuming for simplicity that we are not considering diagonal blocks of $\bar{\mathbf{t}}$ (i.e. $i \neq j, u\neq v$), we have
\begin{align}
\langle &\bar{t}_{(j,i)ba}\bar{t}^*_{(v,u)dc}\rangle =C_{ji} C_{vu}\sum_{p,q=1}^N \langle A^t_{p(j,i)ba}A^{t*}_{q(v,u)dc}\rangle\langle e^{i[(\mathbf{k}_i - \mathbf{k}_j)\cdot\mathbf{r}_p  - (\mathbf{k}_u - \mathbf{k}_v)\cdot\mathbf{r}_q]}\rangle\label{ttcorr}.
\end{align}
The sum in Eq. (\ref{ttcorr}) can be separated into two types of terms: those for which $p=q$ and those for which $p\neq q$. Assuming that the particles in the medium are statistically independent in all senses, the terms for which $p\neq q$ decouple and, in the limit $N \to \infty$, the right hand side of Eq. (\ref{ttcorr}) reduces to the product $\langle \bar{t}_{(j,i)ab}\rangle\langle\bar{t}^{*}_{(v,u)cd}\rangle$. In handling the terms for which $p =q$, the average of the complex exponential can be dealt with as in Eq. (\ref{averaget}). Setting $\bm{\eta} = \mathbf{k}_i - \mathbf{k}_j  - \mathbf{k}_u + \mathbf{k}_v$, we find
\begin{equation}
N\langle e^{i\bm{\eta}\cdot\mathbf{r}}\rangle = (2\pi)^2n\Delta L\sinc\Big(\eta_{z}\frac{\Delta L}{2}\Big)\delta(\eta_x)\delta(\eta_y)\label{uav}.
\end{equation}
The right hand side of Eq. (\ref{uav}) is non-zero when $\eta_x = \eta_y = 0$, i.e. 
\begin{align}
{k}_{ix} - k_{jx} = {k}_{ux} - k_{vx},\qquad\qquad 
{k}_{iy} - k_{jy} = {k}_{uy} - k_{vy}.\label{memory}
\end{align}
This condition is precisely that of the memory effect, which manifests here as a correlation between certain pairs of transmission matrix blocks \cite{tsang2000scattering}. Incorporating this result into Eq. (\ref{ttcorr}), we find that 
\begin{align}
\begin{split}
\langle \bar{t}_{(j,i)ba}\bar{t}^{*}_{(v,u)dc}\rangle &- 	\langle \bar{t}_{(j,i)ba}\rangle \langle\bar{t}^{*}_{(v,u)dc}\rangle\\ 
&= \delta^{R}C_{ijuv}\langle A^t_{(j,i)ba}A^{t*}_{(v,u)dc}\rangle
\sinc\Big(\frac{\Delta L}{2}(k_{iz}-k_{jz}-k_{uz}+k_{vz})\Big)\label{tcovar},
\end{split}
\end{align}
where $C_{ijuv} = wn\Delta L/(k^2\sqrt{k_{iz}k_{jz}k_{uz}k_{vz}})$ and $\delta^R = 1$ when Eq. (\ref{memory}) is satisfied and 0 otherwise. The superscript $R$ here stands for `regular' correlations (to be contrasted with `pseudo' correlations shortly). An analogous result holds for $\langle \bar{r}_{(j,i)ba}\bar{r}^{*}_{(v,u)dc}\rangle - 	\langle \bar{r}_{(j,i)ba}\rangle \langle\bar{r}^{*}_{(v,u)dc}\rangle$, which can be found in Appendix \ref{appendix:a}.

Calculating the correlation in Eq. (\ref{tcovar}) requires knowledge of the scattered field due to a single particle, which is described by the elements of the matrices $\mathbf{A}^{t}_p$ and $\mathbf{A}^{r}_p$. It is worth noting, however, that these correlations can be equivalently described by ensemble averaged Mueller matrices for the slab. Transformations between Mueller matrix elements and field correlations are well documented in the literature (see for example Ref. \cite{perez2017polarized}). While both formalisms are informationally equivalent, reformulating the theory presented here in terms of Mueller matrices may be preferable in some circumstances. For example, decompositions of the Mueller matrix are well known and allow one to express a Mueller matrix in terms of simpler matrices that correspond to familiar optical elements, such as a diattenuator, retarder and depolarizer \cite{Lu:96}. For the purpose of modelling a random medium, it may be simpler to begin with a custom Mueller matrix with desired scattering characteristics, which can then be translated into the corresponding field correlations. Moreover, the Mueller matrix is relatively easy to determine experimentally as it can be calculated from intensity measurements, without requiring interferometric techniques. In cases where the form of a Mueller matrix is known, but analytic expressions for $\mathbf{A}^{t}_{p}$ and $\mathbf{A}^{r}_p$ are not, the Mueller matrix still allows for the extraction of covariances that can used in numerical simulations. 

In addition to regular correlations as in Eq. (\ref{ttcorr}), it is also necessary to consider `pseudo' correlations, i.e. correlations of the form $\langle \bar{t}_{(j,i)ba}\bar{t}_{(v,u)dc}\rangle$ without complex conjugation of the second term. These can be calculated in a similar manner to the regular correlations, yielding the pseudo-covariance
\begin{align}
	\begin{split}
		\langle \bar{t}_{(j,i)ba}\bar{t}_{(v,u)dc}\rangle &- 	\langle \bar{t}_{(j,i)ba}\rangle \langle\bar{t}_{(v,u)dc}\rangle\\ 
		&= \delta^PC_{ijuv}\langle A^t_{(j,i)ba}A^{t}_{(v,u)dc}\rangle
		\sinc\Big(\frac{\Delta L}{2}(k_{iz}-k_{jz}+k_{uz}-k_{vz})\Big)\label{tpseudo},
	\end{split}
\end{align} 
where $\delta^P = 1$ when
\begin{align}
{k}_{ix} - k_{jx} = -({k}_{ux} - k_{vx}),\qquad\qquad 
{k}_{iy} - k_{jy} = -({k}_{uy} - k_{vy}).
\end{align}
and 0 otherwise. It is worth nothing that pseudo-correlations do not influence the statistics of any individual, non-diagonal $2\times 2$ block within the transmission matrix, which can be seen by noting that $\delta^P = 0$ for $i=u$ and $j=v$ ($i \neq j$). Given that non-diagonal blocks also have 0 mean, it follows that every element within a non-diagonal block of the transmission matrix is a circularly symmetric complex random variable. The joint statistics of all of the elements of the transmission matrix, however, do not obey circular symmetry, owing to the presence of pairs of modes for which $\delta^P \neq 0$. For example, consider the pair of blocks $\bar{\mathbf{t}}_{(j,i)}$ and $\bar{\mathbf{t}}_{(i,j)}$, which are related by swapping the incident and outgoing plane wave directions. Referring to Eq. (\ref{t}), the complex exponential terms for these blocks are given by $\exp[i(\mathbf{k}_i - \mathbf{k}_j)\cdot \mathbf{r}_p]$ and $\exp[i(\mathbf{k}_j - \mathbf{k}_i)\cdot \mathbf{r}_p] = \exp[-i(\mathbf{k}_i - \mathbf{k}_j)\cdot \mathbf{r}_p]$ respectively. Thus, regardless of the distribution of the particles within the medium, the complex exponential terms associated with $\bar{\mathbf{t}}_{(i,j)}$ are always the complex conjugates of those associated with $\bar{\mathbf{t}}_{(j,i)}$. This manifests as a non-zero pseudo-correlation between the elements of the matrices $\bar{\mathbf{t}}_{(j,i)}$ and $\bar{\mathbf{t}}_{(i,j)}$, for which it is simple to show that $\delta^P = 1$. Analogously, pseudo-covariances can be found for the other blocks of the scattering matrix, a summary of which is given in Appendix \ref{appendix:a}. 

Finally, we note that correlations (both regular and pseudo) between elements of different blocks of the scattering matrix, e.g.  
$\langle \bar{t}_{(j,i)ba}\bar{r}^{*}_{(v,u)dc}\rangle$, can be computed in an identical fashion to those presented. For simplicity, however, we neglect these so that each of the blocks of the scattering matrix, now assumed to be uncorrelated, can be generated independently. The effects of these additional correlations will be investigated in future works.

\subsection{Matrix symmetries and random matrix generation}\label{symmetries}
In addition to the correlations discussed in the previous section, additional relationships exist between the elements of the scattering matrix due to fundamental physical laws. Provided that there is no absorption or gain within the slab and that the scattering medium satisfies the reciprocity principle, the scattering matrix is constrained to be unitary ($\mathbf{S}^\dagger \mathbf{S} = \mathbb{I}$) and to possess certain lines of symmetries about which some of its elements are equal \cite{PhysRevResearch.3.013129}. These constraints must be satisfied in order for the scattering matrix to represent a physically admissible scattering medium. In order to generate a random scattering matrix that automatically satisfies these symmetry constraints, it is first necessary to identity a set of independent parameters that fully capture the degrees of freedom of the matrix. Once these parameters have been determined, the matrix elements can be uniquely determined from the constraints. Importantly, the set of independent parameters must be chosen so that their statistics can be related to the physical properties of the scattering medium. While it is straightforward to accommodate the reciprocity constraint, unitarity, which manifests as a large system of quadratic equations, is far less trivial to satisfy. A common strategy employed in theoretical studies is the generalized polar decomposition, which parametrizes the scattering matrix in terms of the singular values and vectors of its transmission and reflection matrix blocks \cite{RevModPhys.69.731}. The connection between these parameters and the raw elements of the scattering matrix, however, is non-trivial and unintuitive. Furthermore, the singular vectors still comprise unitary matrices, and thus the problem of how to randomly sample a unitary matrix with given statistics remains.

Instead of directly generating a random unitary matrix, an alternative strategy is to first generate a non-unitary scattering matrix $\mathbf{S}'$ with desired statistical properties and to then find a unitary matrix $\mathbf{S}$ that closely approximates $\mathbf{S}'$. Naturally, the resulting unitary matrix $\mathbf{S}$ from this procedure will not possess the same statistical properties as those prescribed for $\mathbf{S}'$. Provided that the matrix $\mathbf{S}$ is sufficiently `close' to $\mathbf{S}'$ (in the sense that $||\mathbf{S}' - \mathbf{S}||$ is small for some choice of matrix norm), however, this issue becomes unimportant. Given any arbitrary matrix $\mathbf{S}'$, it is well known that the closest unitary approximation $\mathbf{S}$ of $\mathbf{S}'$ is given by the unitary matrix that appears in the polar decomposition of $\mathbf{S}'$ \cite{doi:10.1080/0025570X.1975.11976482}. 

Using the results of Section \ref{statistics}, $\bar{\mathbf{t}}$, $\bar{\mathbf{r}}$ and $\bar{\mathbf{r}}'$ can be generated using a multivariate Gaussian distribution. For diagonal blocks of $\bar{\mathbf{t}}$, as there is no phase variation in Eq. (\ref{t}), the matrix elements are non-random and we can instead use the result for the mean transmission matrix in Eq. (\ref{tavfinal}) as a fixed, non-random value. If reciprocity holds, it is unnecessary to generate $\bar{\mathbf{t}}'$ as it can always be calculated from $\bar{\mathbf{t}}$ (see Ref. \cite{PhysRevResearch.3.013129}). Furthermore, reciprocity of $\bar{\mathbf{r}}$ and $\bar{\mathbf{r}}'$ is automatically enforced by a subset of the correlations in Section \ref{statistics}. Given $\bar{\mathbf{t}}$, $\bar{\mathbf{r}}$, $\bar{\mathbf{t}}'$ and $\bar{\mathbf{r}}'$, which form the non-unitary scattering matrix $\mathbf{S}'$, we then take the unitary part of the polar decomposition of $\mathbf{S}'$ to arrive at a unitary scattering matrix $\mathbf{S}$. Note that in light of, for example, Eq. (\ref{tcovar}), the squared magnitudes of the elements of $\bar{\mathbf{S}}'$ are proportional to the thickness $\Delta L$. In the limit $\Delta L \to 0$, it is clear that $\bar{\mathbf{t}}, \bar{\mathbf{t}}' \to \mathbb{I}$ and   $\bar{\mathbf{r}}, \bar{\mathbf{r}}' \to \mathbb{O}$. The unitary approximation $\mathbf{S}$ also improves in accuracy as $\Delta L$ decreases, satisfying $\lim_{\Delta L\to 0}||\mathbf{S}' -\mathbf{S}|| = 0$.

Given the assumption of single scattering, we may only directly generate scattering matrices for thin slabs. Matrices for slabs of arbitrary thickness, however, can be found by cascading many independent realizations of thin slabs. This is easily achieved using transfer matrices, which possess the useful property that the transfer matrix for a system composed of two contiguous slabs is given by the correctly-ordered product of the transfer matrices of the individual slabs \cite{RevModPhys.69.731}. Scattering matrices can also be cascaded, but the calculation is more complex (see Appendix \ref{appendix:b}). An additional complication however is that the statistical results in Section \ref{statistics} assume that the slab is centred at $z=0$. This led to the emergence of the sinc factors in the expressions for the covariances and pseudo-covariances associated with the matrix elements. If instead the slab were centred at an arbitrary position $z = L_0$, these factors would be different. By performing a change of coordinates, we find that 
\begin{align}
	\bar{\mathbf{S}}^{L_0} = \Lambda_{\pm}^{L_0}\bar{\mathbf{S}}^0\Lambda_{\pm}^{L_0}, \quad\quad \bar{\mathbf{M}}^{L_0} = \Lambda_{\mp}^{L_0}\bar{\mathbf{M}}^0\Lambda_{\pm}^{L_0},
	\label{propagatorEq}
\end{align} 
where $\bar{\mathbf{S}}^{0}$ and $\bar{\mathbf{S}}^{L_0}$ are scattering matrices for the same physical medium, but located with centers at $z=0$ and $L_0$ respectively. The matrices $\bar{\mathbf{M}}^{0}$ and $\bar{\mathbf{M}}^{L_0}$ are the corresponding transfer matrices. The matrices $\Lambda_{\pm}^{L_0}$ and $\Lambda_{\mp}^{L_0}$ are diagonal matrices containing complex phasor terms, more details of which can be found in Appendix C. Thus, in order to generate a random matrix describing a scattering medium centred at $z = L_0$, we can first generate $\bar{\mathbf{S}}^0$, whose statistics are given by the results of Section \ref{statistics}, and then compute $\bar{\mathbf{S}}^{L_0}$ using Eq. (\ref{propagatorEq}). 

Consider now the special case of a series of slabs, all of equal thickness $\Delta L$, positioned contiguously in the $z$ direction so as to constitute a single, continuous medium. In this case, in order to find the scattering or transfer matrix for the overall medium, it can be shown that it is sufficient to take the product of transfer matrices of the form $\bar{\mathbf{M}} = \Lambda_{\pm}^{\Delta L}\bar{\mathbf{M}}^{0}$, where $\bar{\mathbf{M}}^{0}$ can be randomly generated using the statistics in Section \ref{statistics} and the method outlined in this section. Taking the product of $N$ such transfer matrices yields a transfer matrix for a scattering medium of thickness $N\Delta L$. More details can be found in Appendix \ref{appendix:propagator}. Finally, if necessary, scattering at the boundaries of the slab can also be incorporated into the matrix cascade by including additional scattering or transfer matrices that capture the surface effects at either interface.

\section{Numerical simulations}\label{simulations}
In this section we discuss numerical simulations of scattering matrices, performed for random media containing different types of particles. We first outline our simulation method and then present some results with discussion.
\subsection{Method}\label{method}
Before generating random scattering matrices, it is first necessary to choose a set $K$ of transverse wavevectors and associated weights. Since we need only consider homogeneous plane waves, the set of all possible transverse wavevectors in $k$-space is the interior of the circle $|\bm\kappa|^2 = k_x^2 + k_y^2 = k^2$. In a real scattering experiment, the number of independent modes can be extremely large, on the order of millions per square millimetre of illuminated surface area \cite{Mosk2012}. In our simulations, however, there is a practical upper limit to the number of modes that can be used, as large scattering matrices quickly become unwieldy and computationally intensive. We distributed modes on a Cartesian grid in $k$ space, including the origin and with lattice spacing given by $\Delta k_x = \Delta k_y = 0.1715k$, rejecting modes lying on lattice points for which $k_x^2 + k_y^2 > k^2$. This spacing was chosen arbitrarily so that the set $K$ contains a total of $101$ modes, which, given the block structure of $\bar{\mathbf{S}}$, means our scattering matrices were of size $404\times 404$. Of course, as the boundary of $k$ space is a circle, the interior cannot be fully tessellated by a Cartesian grid and modes close to the boundary have associated weights not given by the product $\Delta k_x \Delta k_y$. To ensure that the weight for each mode was equal and that the weights were correctly normalized, we decided to give each mode the weight $w = \pi k^2/101$. This value differs slightly to $\Delta k_x \Delta k_y$, but this discrepancy decreases as the number of modes increases.

We simulated two types of scattering media: one containing spherical, optically inactive particles and one containing chiral particles exhibiting circular birefringence. In either case, the single particle scattering properties are known theoretically (see, for example, Ref. \cite{bohren2008absorption}). It was convenient to specify the matrix $\mathbf{A}^{t/r/r'}_{(j,i)}$ as a product of rotation matrices and a $2 \times 2$ scattering matrix defined with respect to the scattering plane. Details of this calculation can be found in Appendix \ref{appendix:rotation}. We chose the wavelength $\lambda = 500\,\mathrm{nm}$ and considered isotropic spheres of three different sizes, namely $x = 1,2$ and $4$, where $x = ka$ is the dimensionless size parameter and $a$ is the particle radius. For each particle size, we used the same relative refractive index $m=1.2$ and calculated the $\mathbf{A}^{t/r/r'}_{(j,i)}$ factors using Mie theory. In addition, we performed simulations for chiral spheres of two different size parameters, $x=1$ and 4. For both size parameters, we chose a mean relative refractive index $\bar{m} =1.2$ and circular birefringence $\Delta m = 0.044$ so that $m_l = \bar{m} + \Delta m$ and $m_r = \bar{m} - \Delta m$ were the relative refractive indices experienced by incident left and right handed circular polarization respectively. This birefringence is such that left handed circularly polarized light is more strongly forward scattered than right handed circularly polarized light. 

For a given type of particle, the volume density $n$ and slab thickness $\Delta L$ that appear in the expressions for the mean and covariances in Section \ref{statistics} are free parameters, not immediately constrained by any other variables. It is important, however, that these parameters are chosen in a way that does not violate any of the basic assumptions made in our model. To ensure that this was the case, we identified three conditions that must be simultaneously satisfied. Firstly, we require $kd \gg 1$, where $d = (1/n)^{1/3}$ is a measure of the average spacing between the particles in the medium. This condition ensures that the particles are all in the far field of each other. Secondly, we require $l/\Delta L \gg 1$, where $l$ is the mean free path of medium, given by the standard formula $l = (n\sigma)^{-1}$, where $\sigma$ is the scattering cross section \cite{bohren2008absorption}. This condition ensures that the single scattering approximation holds. Since this second condition requires that the slab thickness $\Delta L$ is small, we identified a third condition $\Delta L/2a > 1$ that ensures that the slab is thick enough to contain the particles. 

Instead of specifying $n$ directly, it was simpler to start with a particle volume fraction $\phi$ and calculate the density via $n = \phi/V_p$, where $V_p$ is the volume of a single particle. For all simulations we chose the value $\phi = 0.01$. In specifying $\Delta L$, a problem we encountered was that, given the appearance of $1/k_{iz}$ factors in, for example, Eq. (\ref{tavfinal}), the numerical values of the means and covariances can become large for grazing incidence modes. In effect, these modes `see' a larger thickness for the scattering medium. To overcome this problem, we set a threshold value $\delta \ll 1$ and demanded that the elements of the mean transmission matrix were smaller than $\delta$ for all incident modes (i.e. all blocks on the diagonal of $\bar{\mathbf{t}}$). Specifically, for all $i$, we solved the equation $\delta = 2\pi n \Delta L s_{\mathrm{max},i}/(kk_{iz})$ for $\Delta L$, where $s_{\mathrm{max},i}$ is the largest singular value of $\langle\mathbf{A}^t_{(i,i)}\rangle$ and took $\Delta L$ to be the minimum of all these values. We found that using a threshold value $\delta = 0.1$ gave values of $\Delta L$ that satisfied our conditions. A summary of all the simulation parameters is given in Table~\ref{table}, where each row corresponds to a different parameter set. For chiral particles, the presented mean free path is that calculated from Mie theory for an isotropic sphere with the same size parameter.
\begin{table}[]
	\caption{\label{table}Summary of the physical parameters used in simulations.}
	\centering\begin{tabular}{
			c 
			c 
			d{2} 
			c 
			c 
			d{2} 
			d{2} 
			c 
			d{2} 
			c 
			d{1} 
		}
		\toprule
		\multicolumn{3}{c}{Input}
		& \multicolumn{5}{c}{Calculated Parameters}
		& \multicolumn{3}{c}{Physical Checks} \\ \cmidrule(lr){1-3} \cmidrule(lr){4-8} \cmidrule(lr){9-11}
		$x$ 
		& $\overline{m}$
		& \multicolumn{1}{c}{$\Delta m$}
		& \multicolumn{1}{c}{$n/ \mu \mathrm{m}^{-3}$}
		& \multicolumn{1}{c}{$\Delta L/\mu \mathrm{m}$}
		& \multicolumn{1}{c}{$l/\mu \mathrm{m}$}
		& \multicolumn{1}{c}{$a/\mathrm{nm}$}
		& \multicolumn{1}{c}{$d/\mu \mathrm{m}$}
		& \multicolumn{1}{c}{$kd$}
		& \multicolumn{1}{c}{$\Delta L/2a$}
		& \multicolumn{1}{c}{$l/\Delta L$}
		\\
		\midrule
		1 & 1.2 & 0 	& 4.737 & 1.177 & 311.57 & 79.58 & 0.595 & 7.48 & 7.34 & 264.7 \\
		2 & 1.2 & 0 	& 0.592 & 1.126 & 88.08 & 159.15 & 1.191 & 14.96 & 3.53 & 78.2 \\
		4 & 1.2 & 0 	& 0.074 & 1.173 & 35.87 & 318.31 & 2.382 & 29.93 & 1.84 & 30.6 \\ 
		1 & 1.2 & 0.044 & 4.737 & 0.969 & 311.57 & 79.58 & 0.595 & 7.48 & 6.09 & 321.7 \\
		4 & 1.2 & 0.044 & 0.074 & 0.969 & 35.87 & 318.31 & 2.382 & 29.93 & 1.52 & 37.0 \\ \bottomrule
		\end{tabular}

		\end{table}

For each parameter set we generated the matrices $\bar{\mathbf{t}}$, $\bar{\mathbf{r}}$ and $\bar{\mathbf{r}}'$ using a multivariate Gaussian distribution, calculating $\bar{\mathbf{t}}'$ from $\bar{\mathbf{t}}$ as previously discussed. For each matrix $\bar{\mathbf{S}}'$ we computed the unitary approximation $\bar{\mathbf{S}}$ as described in Section \ref{symmetries} and its associated transfer matrix $\bar{\mathbf{M}}$. To properly account for propagation along the $z$ axis when cascading multiple slabs, we then pre-multiplied each of these transfer matrices by the constant matrix $\Lambda_{\pm}^{\Delta L}$. In total, we randomly generated pools of $10^4$ transfer matrices for each parameter set for slabs with thicknesses as shown in Table~\ref{table}.
 
In order to access the multiple scattering regime, it is necessary to cascade at least $\sim l/\Delta L$ transfer matrices, which, as can be seen, can be on the order of $10^2$ matrices. Additionally, in order to compute good statistics, it is necessary to have a large number of scattering matrices at any given thickness. Consequently, in total, a large number of random matrices are required to generate data for random media with thicknesses beyond a mean free path. To alleviate this computational burden, we first decided upon a thickness step size ($0.5l$ in our simulations) and calculated a secondary pool of $10^4$ transfer matrices by cascading random selections of transfer matrices from the initial matrix pool so that each resulting transfer matrix corresponded to a random medium of thickness equal to the step size. In generating this secondary pool, some matrices from the initial matrix pool are reused, which may introduce unwanted statistical correlations between members of the secondary pool. Given that the number of possible permutations in performing the matrix products is far greater than any realistic size for the secondary pool, however, we found this issue to be unimportant. Finally, we used an additional set of $10^4$ transfer matrices for actual data collection. For this final set of transfer matrices, we progressed through media of increasing thicknesses in steps of $0.5l$ to a final thickness of $30l$, collecting data at each step. Progressing to the next thickness is performed by multiplying each matrix in our final collection with a randomly selected matrix from the secondary pool. Therefore, after the secondary pool has been generated, no further random matrices are required.

When continuing to multiplying transfer matrices together, the elements tend to diverge, as the set of transfer matrices is not a compact group \cite{PhysRevE.75.031113}. Therefore, after a certain point, it is necessary to convert all matrices used in the calculations into their corresponding scattering matrices. While slower to cascade, unitarity of the scattering matrices means they do not suffer from the same numerical problem.
 
\subsection{Model validation}\label{results}
In the following section we present a variety of statistical data calculated from our simulations for thicknesses $L$ ranging from the single to multiple scattering regimes. As we have access to the entire scattering matrix, in addition to analyzing more familiar characteristics of the scattered field in individual modes, such as the intensity and DoP, we may also calculate parameters that are functions of larger sections of $\bar{\mathbf{S}}$, such as correlations between different blocks or the transmission eigenvalues. In all of the following data, averages were computed over all $10^4$ realizations of the scattering matrix for each thickness.
\subsubsection{Isotropic spheres}
\begin{figure}[t!]
	\centering\includegraphics[width =\textwidth]{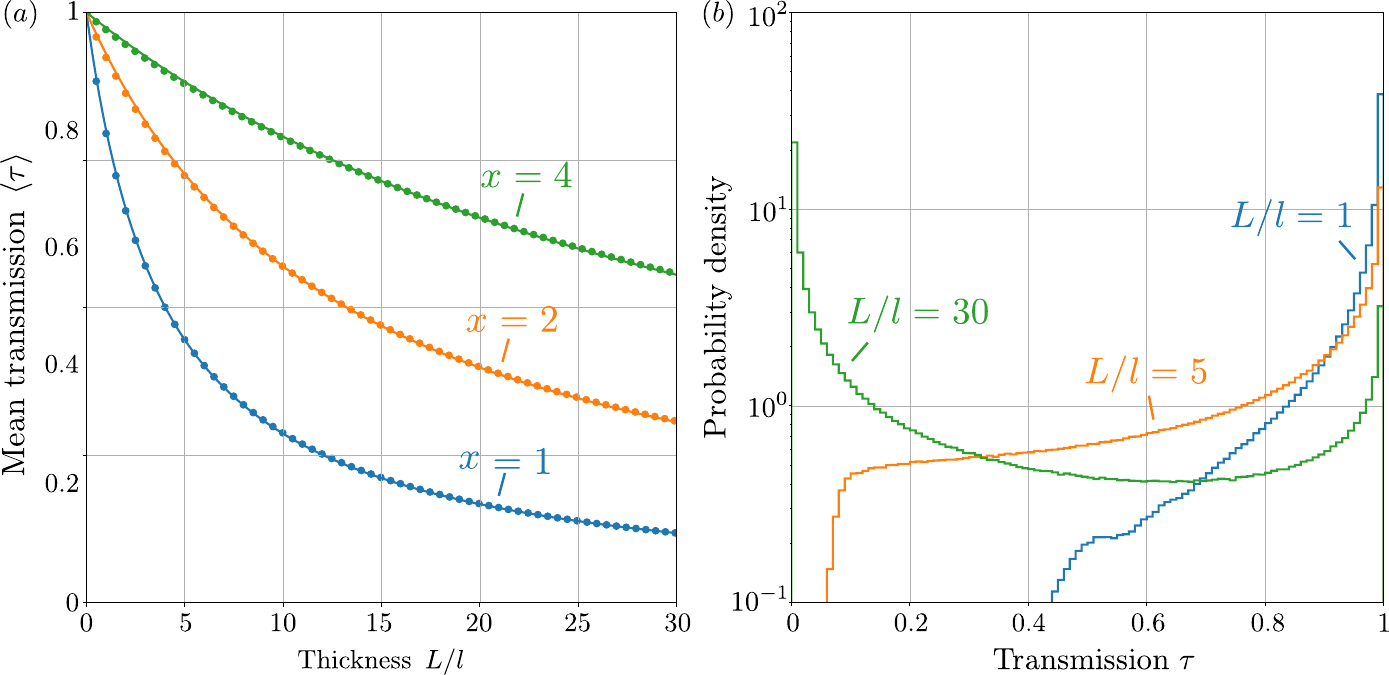}
	\caption{(a) Mean transmission as a function of thickness for size parameters $x=1, 2$ and 4. Fitting curves are of the form $\langle \tau \rangle = (1+L/\alpha l)^{-1}$, where $\alpha$ was calculated from the data points. (b) Probability density functions of transmission eigenvalues for thicknesses $L/l = 1, 5$ and 30 for size parameter $x=2$.  }\label{transmission_eigs}
\end{figure}
The following results are for optically inactive spheres whose parameters are given in the first three rows of Table \ref{table}.

\subsubsubsection{Transmission eigenvalues}\label{intenssection}

Figure \ref{transmission_eigs}(a) shows the mean transmission eigenvalue $\langle \tau \rangle = \langle \mathrm{tr}(\bar{\mathbf{t}}^{\dagger}\bar{\mathbf{t}})\rangle/N$, where $\mathrm{tr}$ denotes the trace operator and $N$ is the size of the transmission matrix, as a function of medium thickness. When all incident light is transmitted, regardless of incident mode or polarization state, $\langle \tau \rangle = 1$, whereas $\langle \tau \rangle$ = 0 when no light is transmitted. By conservation of energy, a decrease in $\langle \tau \rangle$ must be compensated for by an increase in the mean reflection eigenvalue $\langle \rho \rangle= 1 - \langle \tau \rangle = \langle \mathrm{tr}(\bar{\mathbf{r}}^{\dagger} \bar{\mathbf{r}})\rangle /N$. The main characteristics of Figure \ref{transmission_eigs} are that $\langle \tau \rangle$ decreases monotonically with increasing medium thickness, as is known to occur for isotropic systems \cite{RevModPhys.89.015005}, and that the rate of decrease is smaller for larger size parameters. The dependence on particle size can be explained by single particle scattering anisotropy: larger particles preferentially scatter light in the forward direction, which results in a smaller decay rate for $\langle \tau \rangle$. In Ref. \cite{PhysRevB.44.3559}, it was found that in a quasi-one dimensional system with isotropic scattering, to lowest order, the mean transmission eigenvalue decays as $\langle \tau \rangle = (1+L/l)^{-1}$. We found that our curves were reasonably well fit by functions of the form $\langle \tau \rangle = (1+L/\alpha l)^{-1}$, where $\alpha$ is a fitting parameter given by 4.02, 13.25 and 37.51 for $x=1, 2$ and $4$ respectively. Physically, $\alpha l$ can be interpreted as a length scale over which the random medium scatters isotropically, similar to the transport mean free path $l^* = l/(1-g)$, where $g$ is the anisotropy factor \cite{brosseau1998fundamentals}. We found however that our value for $\alpha$ was larger than $1/(1-g)$. To explain this, we note that the expression $1/(1-g)$ only accounts for randomization of direction, whereas $\alpha$ also incorporates isotropization of polarization state. 

Figure \ref{transmission_eigs}(b) shows the probability density function for the transmission eigenvalues of scattering matrices at thicknesses $L/l =1, 5$ and 30 for size parameter $x=2$. The distribution transitions from being highly peaked at $\tau =1$ for small thicknesses to highly peaked at $\tau = 0$ for large thicknesses. Notably, even for the largest thickness $L/l = 30$, there still exist channels for which $\tau =1$. These open eigenchannels are well known and have been studied extensively, both theoretically and experimentally, particularly for scalar waves \cite{RevModPhys.89.015005, Mosk2012}. In our simulations however, these eigenchannels also have a specific polarization structure. In order to construct such an eigenchannel experimentally, such as in a wavefront shaping experiment, it would be necessary to control both the relative intensity and polarization state of each plane wave component of the incident field. Considering the eigenchannel with largest transmission, we found that altering the polarization state of any individual plane wave component while keeping its relative intensity constant resulted in a decrease of the total transmitted intensity. Careful control of the incident polarization state may therefore lead to enhanced transmission over the case of scalar waves. We found similar behaviour for $x=1$ and 4, but the rate at which the distribution evolves with thickness is greater for $x=1$ and smaller for $x=4$, as expected due to scattering anisotropy.

\begin{figure}[t!]
	\centering\includegraphics[width =1\textwidth]{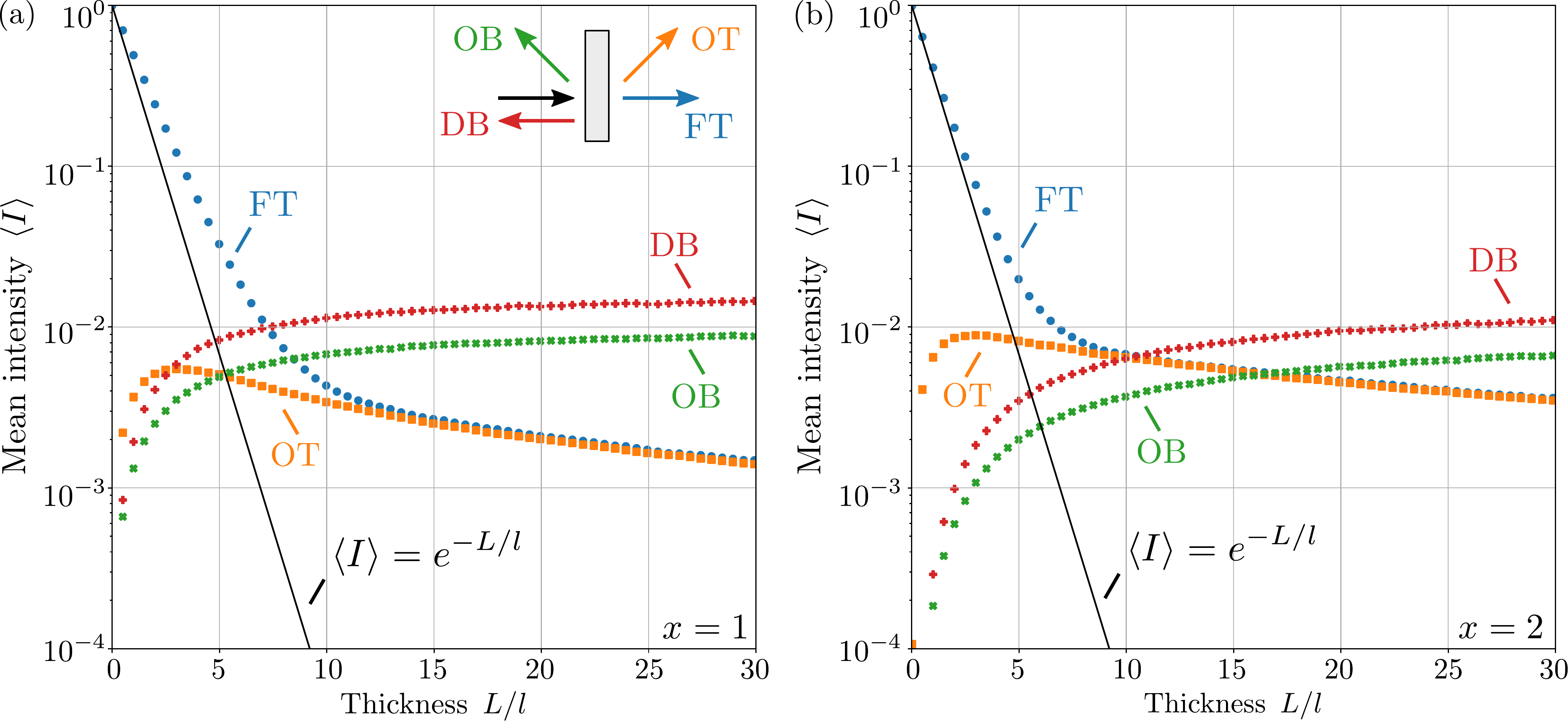}
	\caption{Mean intensity as a function of thickness for size parameters (a) $x=1$ and (b) $x=2$. The intensity is shown in four different outgoing modes: forward transmission (FT), oblique transmission (OT), oblique backscattering (OB) and direct backscattering (DB). A visual aid is provided in (a).}\label{mean_intensity_mie}
\end{figure}

\subsubsubsection{Scattered intensity}\label{intenssection}

Figures \ref{mean_intensity_mie}(a) and (b) show the mean plane wave intensity $\langle I \rangle $ in several outgoing modes for a normally incident plane wave and size parameters $x=1$ and $2$. We focused our attention on four different modes: the transmitted wave parallel to the incident field (forward transmission, or FT); the transmitted wave for which $\bm \kappa/k \approx (3\Delta k_x, 0)^\mathrm{T}$ (oblique transmission, or OT); the reflected wave for which $\bm \kappa/k = (3\Delta k_x, 0)^\mathrm{T}$ (oblique backscattering, or OB) and the backscattered wave propagating in the opposite direction to the incident field (direct backscattering, or DB). For each mode, $\langle I \rangle$ was calculated by taking the ensemble average vector norm of the first column of the appropriate matrix block. Since the scatterers are isotropic, $\langle I \rangle$ is independent of incident polarization state. 

Observing FT in Figure \ref{mean_intensity_mie}(a), we see that $\langle I \rangle$ decays exponentially, but the decay rate changes at around $L/l \sim 10$, becoming smaller for large thicknesses. The initial exponential decay is the well-known Beer-Lambert law, which is given by $\langle I \rangle = e^{-L/l}$ and is shown in the figure as a black line. For larger thicknesses, the change in decay rate occurs due to light being scattered back into the forward direction (i.e. an increase in the `incoherent' intensity). The notable bend in the decay curve can therefore be thought of as a transition to the multiple scattering regime. Before this transition occurs, our data points are systematically larger than those predicted by the Beer Lambert law, which we attribute to numerical inaccuracies stemming from our simplistic cubature scheme.
 
Looking at OT in Figure \ref{mean_intensity_mie}(a), for small thicknesses we see that the intensity is small and increases with thickness. In this regime, scattering is weak and intensity increases as more light is scattered away from FT and into OT. For large thicknesses, the intensity behaviour is similar to FT, settling on a limiting decay trajectory. The behaviour in reflection is conjugate to that of transmission. In OB, the intensity is initially small, but increases monotonically. The same behaviour is observed in DB, but the intensity values are $\sim 1.8$ times larger. This intensity enhancement is a signature of the coherent backscattering effect, which emerges naturally from our simulations from the enforcement of reciprocity in the scattering matrices. This enhancement is less than ideal (a factor of 2) due to the non-zero size each mode occupies in $k$-space. Figure \ref{mean_intensity_mie}(b) shows similar trends to Figure \ref{mean_intensity_mie}(a). The most notable differences are that the reflected intensities increase at slower rates and the transmitted intensities decay at a slower rate, both of which are also a result of scattering anisotropy.

\subsubsubsection{Degree of polarization}\label{miedop}

\begin{figure}[t!]
	\centering\includegraphics[width =\textwidth]{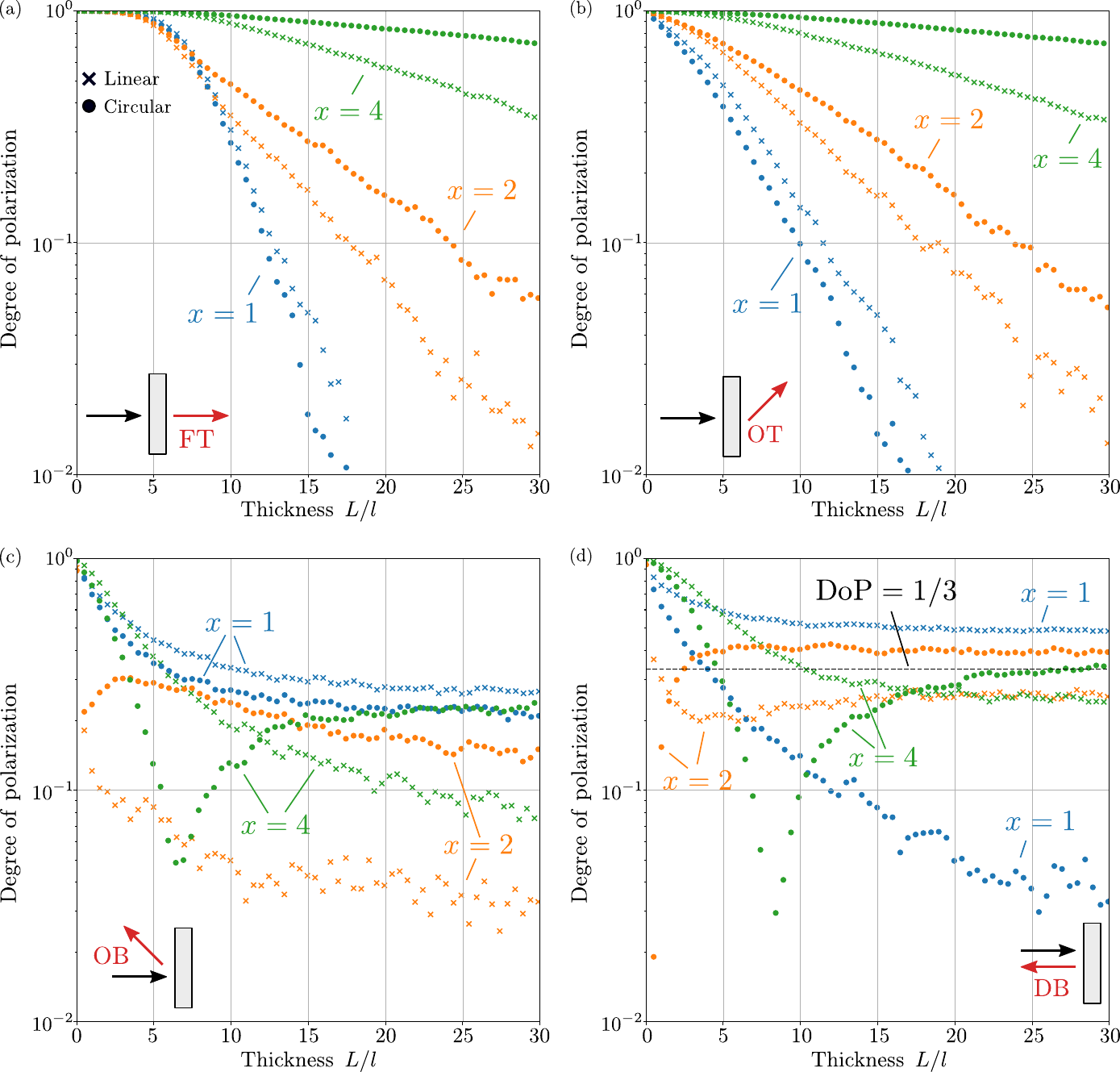}
	\caption{Degree of polarization as a function of thickness for incident linearly ($\times$ markers) and circularly ($\circ$ markers) polarized light and size parameters $x=1$ (blue), 2 (orange) and 4 (green) in (a) forward transmission (FT), (b) oblique transmission (OT), (c) oblique backscattering (OB) and (d) direct backscattering (DB).}\label{dopgraphs}
\end{figure}

In Figure \ref{dopgraphs}, we show the DoP in the same four modes discussed in Section \ref{intenssection} for both a linearly and circularly polarized, normally incident plane wave. The DoP can be found by calculating the ensemble average Mueller matrix for each mode, from which the average scattered Stokes vector for different incident polarization states, and thus the DoP, can be deduced. We emphasize that for any individual realization of a scattering medium the scattered field is fully polarized. The DoP in this context is therefore a measure of the distribution of scattered polarization states across the ensemble of random media. 

Figure \ref{dopgraphs}(a) shows the DoP versus thickness in FT. As is evident from the graph, the DoP decays more slowly for larger particles, regardless of the incident polarization state. Furthermore, for $x=1$, we see that linear polarization better preserves its DoP over greater thicknesses than circular polarization, but the opposite is true for $x=2$ and 4. This phenomenon, sometimes called the polarization memory effect, is well understood and can be explained by scattering anisotropy \cite{brosseau1998fundamentals, PhysRevE.72.065601}. A similar trend can be observed in Figure \ref{dopgraphs}(b), which shows the DoP in OT. The most notable difference is that, particularly for $x=1$, the DoP begins to decay immediately, as opposed to at $L/l \sim 5$ for FT. This is due to the presence of the incident field in FT and absence thereof in OT.

The behaviour of the DoP in OB, as shown in Figure \ref{dopgraphs}(c) is much more interesting. The most obvious feature is that the DoP retains a residual, non-zero value as $L/l \to \infty$ for all particle sizes and polarization states. This residual DoP can be explained by noting that in reflection, unlike transmission, a significant contribution to the total field comes from low-order scattering sequences that occur close to the medium’s surface \cite{Zimnyakov_2001}. Another striking feature is the non-monotonicity of the DoP for circular polarization and size parameters $x=2,4$ (and the absence of such behaviour for $x=1$). Specifically, the DoP can be seen to dip to a minimum value before increasing again and settling on a limiting value. This occurs at $L/l \sim 0.5$ for $x=2$ and at $L/l \sim 6.5$ for $x=4$. There is also a non-trivial dependence between the limiting DoP value, size parameter and incident polarization state.

To explain some of these phenomena, we note that, roughly speaking, the reflected field is the sum of three types of contributions: low scattering order contributions from scattering sequences occurring close to the medium's surface (type I); polarization-randomizing, high order scattering contributions from long, circuitous sequences deep within the medium (type II) and polarization-maintaining, high order scattering contributions from long, largely forward-directed sequences deep within the medium (type III). As type I contributions occur near the slab boundary, their overall magnitude should be largely independent of thickness. The latter two contributions, however, should increase in magnitude with thickness. 

For $x=1$, since large angle scattering is more probable than for $x=2$ or 4, type I contributions dominate the total backscattered field for all thicknesses. The DoP decays relatively slowly as type II contributions, which give a polarization-randomizing background, gradually increase with thickness. As scattering is relatively isotropic, type III contributions are comparatively weak and thus less relevant. To verify this claim, we observed distributions of scattered polarization states over the Poincaré sphere for different thicknesses. We found that for all thicknesses, these distributions remained concentrated at the polarization state that would result from a single backscattering event, with an increasing isotropic background for larger thicknesses. 

The situation is different for $x= 2$ and 4. Since larger particles scatter more strongly in the forward direction, type I contributions, which require large angle scattering events, are comparatively much weaker. For incident linearly polarized light, type I and III contributions both tend to preserve the incident polarization state. Although type I contributions are weaker for $x=4$ than $x=2$, type III contributions are greater for $x=4$ than $x=2$. There is thus a non-trivial relationship between the relative magnitudes of these contributions as particle size changes, the exact balance of which dictates the non-monotonicity of the limiting value of the DoP for linear polarization. 

For $x=2$ and $4$, the situation is again different for incident circularly polarized light. While type III contributions maintain incident helicity, type I contributions result in a helicity flip. Therefore, in transitioning from small to large thicknesses, the distribution of scattered states on the Poincaré sphere must transition from being highly focused at the helicity flipped pole (a single scattering, type I dominant regime) to being relatively isotropic, but concentrated at the pole with the same helicity as the incident field (a multiple scattering, type III dominant regime). Although both of these extremes correspond to relatively large values for the DoP, in performing this transition, there is an intermediate thickness at which the distribution of scattered states on the Poincaré sphere shows no preference for either pole, in which case the DoP is small. It is precisely this thickness that corresponds to the dips in the DoP. The dip is more obvious for $x=4$ than $x=2$ and occurs at a larger thickness because photons are able to penetrate further into the medium for $x=4$ before their directions are randomized. This behaviour has been observed experimentally in oblique backscattering from suspensions of polystyrene spheres \cite{Sun_2013}.

As a final remark, we note that in Figure \ref{dopgraphs}(d), which shows similar trends to Figure \ref{dopgraphs}(c), the DoP tends to values close to $1/3$ for $x=2$ and 4. This is the value predicted for scattering matrices drawn from the circular orthogonal ensemble in the direct backscattering direction \cite{BYRNES2022127462}. For $x=1$, the dominance of type I contributions to the reflected field means that the phase function of the slab better resembles that of the individual particles in the medium, which is not isotropic. This may explain why the DoP for $x=1$ deviates strongly from this value, particularly for incident circularly polarized light. The assumption of isotropic scattering, which is necessary for the circular ensemble to be an appropriate model, is better satisfied at large thicknesses for $x=2$ and $4$, whose scattered fields are dominated by multiply scattered light.

\subsubsubsection{Diattenuation and retardance}

\begin{figure}[t!]
	\centering\includegraphics[width =\textwidth]{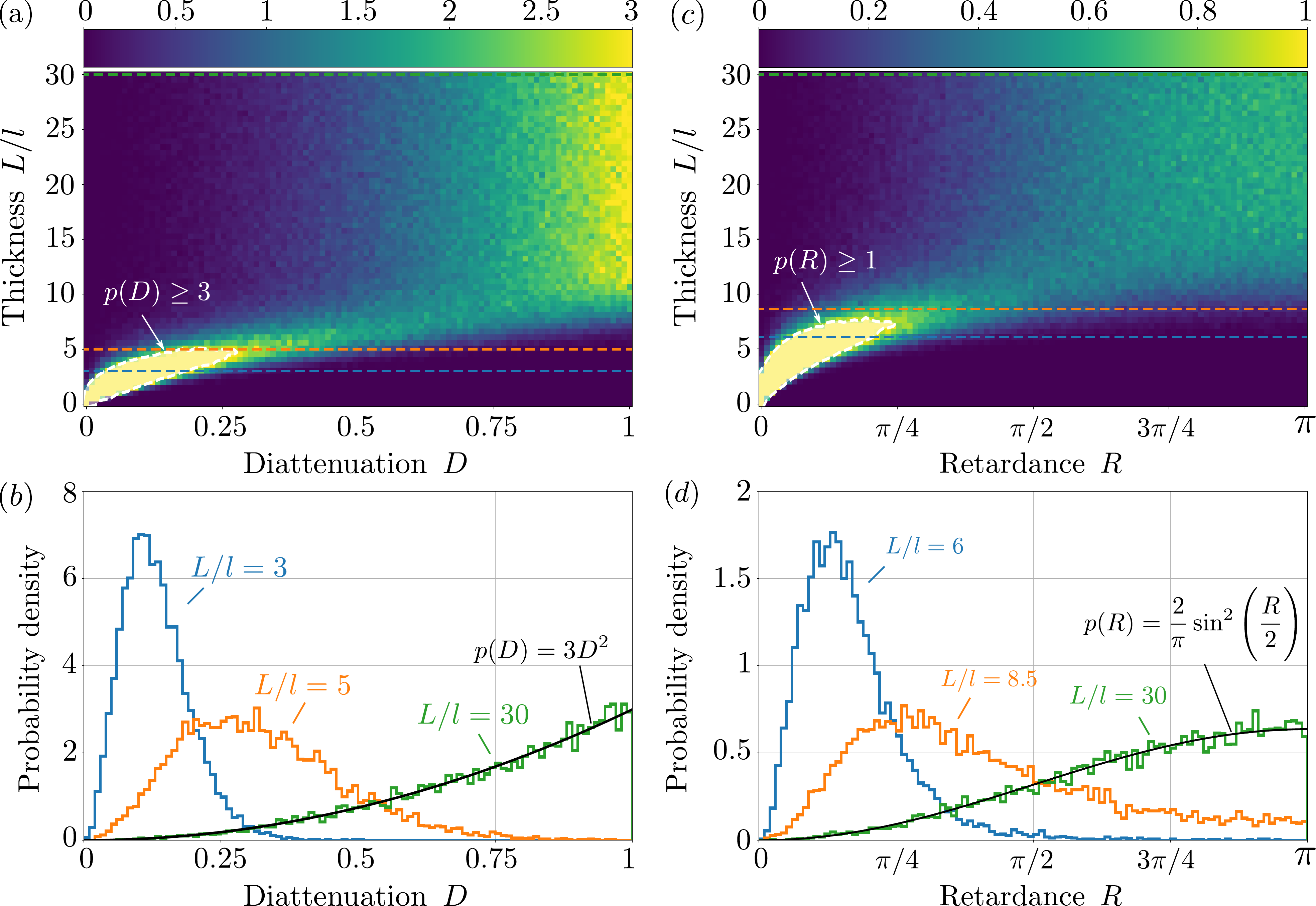}
	\caption{Diattenuation and retardance histograms for size parameter $x=1$ in forward transmission. (a) shows a heatmap of probability distribution functions for diattenuation at different thicknesses. The dashed contour close to the origin indicates a region in which the colors are saturated and the probability density is greater than 3. (b) shows a selection of histograms corresponding to horizontal cross-sections of data in (a). (c) and (d) show analogous data for retardance, with the dashed contour in (c) showing a region for which the probability density is greater than 1.}\label{x=1}
\end{figure}

An additional pair of parameters that can be useful in assessing the polarimetric properties of a scattering medium are diattenuation and retardance. As we have access to the full scattering matrix, these can computed for any $2\times 2$ block using the polar decomposition \cite{perez2017polarized}. Unlike the DoP, which is dependent on the incident polarization state, diattenuation and retardance are computed from the entire $2\times 2$ block. We note that, as the scattering matrix is unitary, the diattenuation we compute is solely due to scattering and not absorption (dichroism).

\begin{figure}[t!]
	\centering\includegraphics[width =\textwidth]{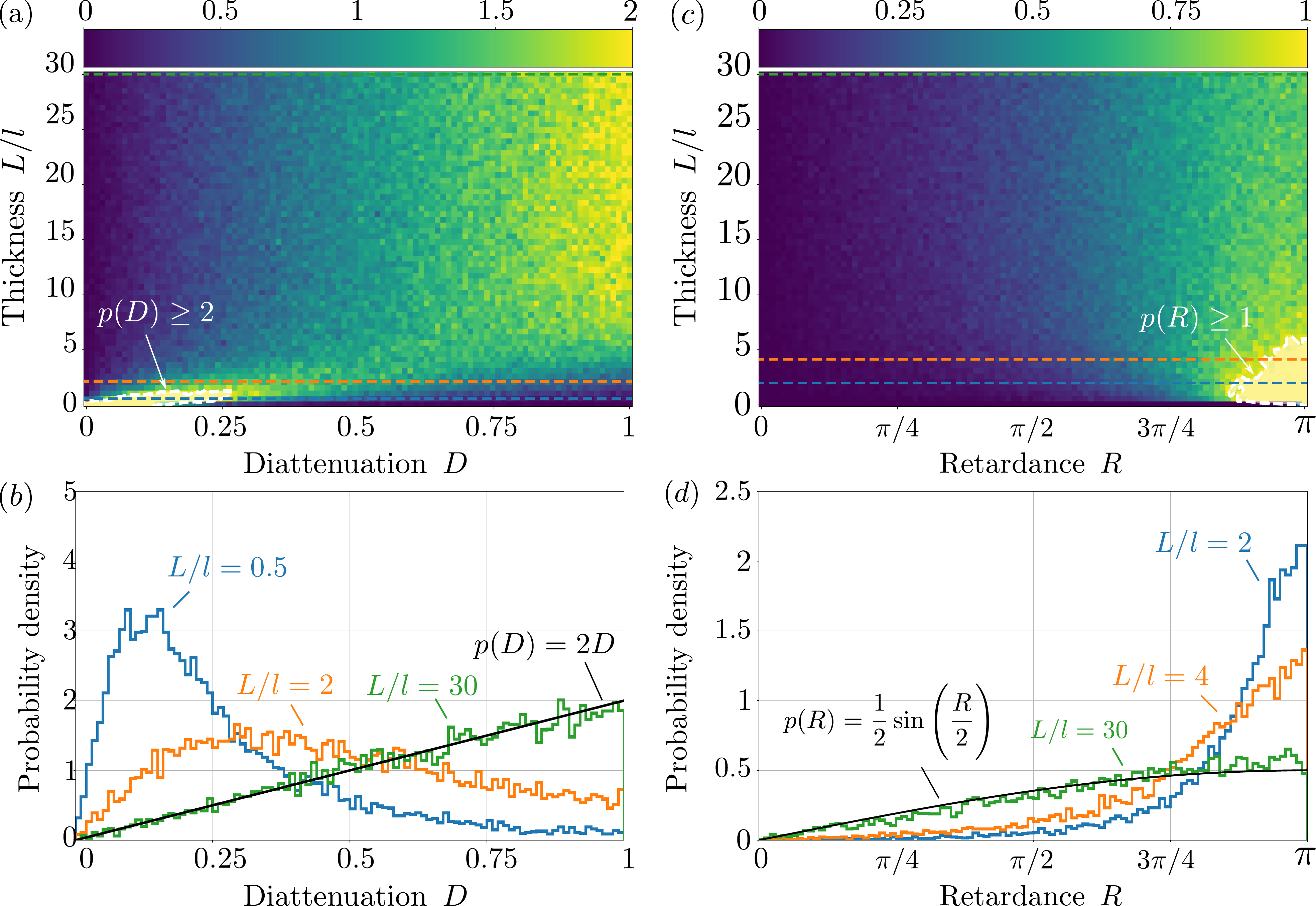}
	\caption{As per Figure \ref{x=1}, albeit for scatterers with size parameter $x=4$ and direct backscattering (DB). Dashed contours in (a) and (c) demarcate regions for which the heat map has been clipped for probability densities $\geq 2$ and $\geq 1$ respectively.}\label{x=4}
\end{figure}

Figure \ref{x=1}(a) shows a heat map of probability density functions for diattenuation $D$ in FT at different thicknesses for $x=1$. The values of the color bar are dimensionless and represent probability density. The color bar values are accurate for thicknesses beyond $10l$, but are saturated for shorter thicknesses in a small region close to the origin as outlined by the dashed contour. In this region, $D$ is strongly peaked close to 0, as a weakly scattering medium, which largely preserves the incident field, cannot be strongly diattenuating. In Figure \ref{x=1}(b), density functions for a selection of thicknesses as indicated by the horizontal dashed lanes in Figure \ref{x=1}(a) are shown more clearly. As can be seen, the diattenuation density function transitions from being a delta function $p(D) = \delta(D)$ at $L=0$ to a limiting distribution given by $p(D) = 3D^2$ as $L \to \infty$. This limiting distribution is precisely that predicted by a random $2\times 2$ matrix of uncorrelated, complex Gaussian entries \cite{BYRNES2022127462}. The transition of the diattenuation distribution is therefore related to the decorrelation of the elements of the scattering matrix. Figures \ref{x=1}(c) and \ref{x=1}(d) show analogous data for retardance in FT. Qualitatively, the behaviour is similar to diattenuation and the density function makes a similar transition from $p(R) = \delta(R)$ to the limiting distribution $p(R) = 2\sin^2(R/2)/\pi$, which is also that predicted by a random Gaussian matrix. For small thicknesses, we found that the distributions of the diattenuation and retardance vectors were concentrated at polarization states expected from single scattering theory. These distributions however became isotropic over the Poincaré sphere for large thicknesses, meaning that no particular polarization state is preferentially scattered on average in the large thickness limit. For individual medium realizations, however, as diattenuation tends to be quite large ($\langle{D}\rangle = 0.75$), there will exist random polarization states that are transmitted much more strongly than others.

Figure \ref{x=4} shows a similar set of plots to those of Figure \ref{x=1}, but for particle size $x=4$ and for DB. The main differences between Figures \ref{x=1} and \ref{x=4} are the behaviour of retardance, the rates of evolution of the density functions and the limiting probability density functions. As shown in Figures \ref{x=4}(a) and \ref{x=4}(b), owing to the absence of the incident field, the diattenuation distribution tends to a limiting distribution (this time given by $p(D) = 2D$) at a shorter thickness. In Figure \ref{x=4}(c), for small thicknesses, the retardance is peaked close to $R = \pi$, which is the value expected by single particle backscattering. The retardance distribution evolves to $p(R) = \sin(R/2)/2$ at larger thicknesses, as can be seen in Figure \ref{x=4}(d). The fact that these limiting densities differ to those in Figure \ref{x=1} is another peculiarity of the DB direction. Due to reciprocity, additional correlations exist between the elements of the $2\times2$ block, even in the large thickness limit. The previous results relevant to a matrix of uncorrelated Gaussian entries therefore no longer apply. It has been shown, however, that these limiting densities are in fact those predicted for diagonal blocks of a random matrix sampled from the circular orthogonal ensemble \cite{BYRNES2022127462}.

\subsubsection{Chiral spheres}
The following results are for chiral spheres, whose parameter sets are given in the final two rows of Table \ref{table}. For these particles, since the mean free path depends on the incident polarization state, to better illustrate the polarization dependence of the statistics of the scattered field we decided to normalize the medium thickness $L$ by the mean free path calculated for an optically inactive sphere with the same size parameter. 
\subsubsubsection{Transmission and reflection}

\begin{figure}[t!]
	\centering\includegraphics[width =\textwidth]{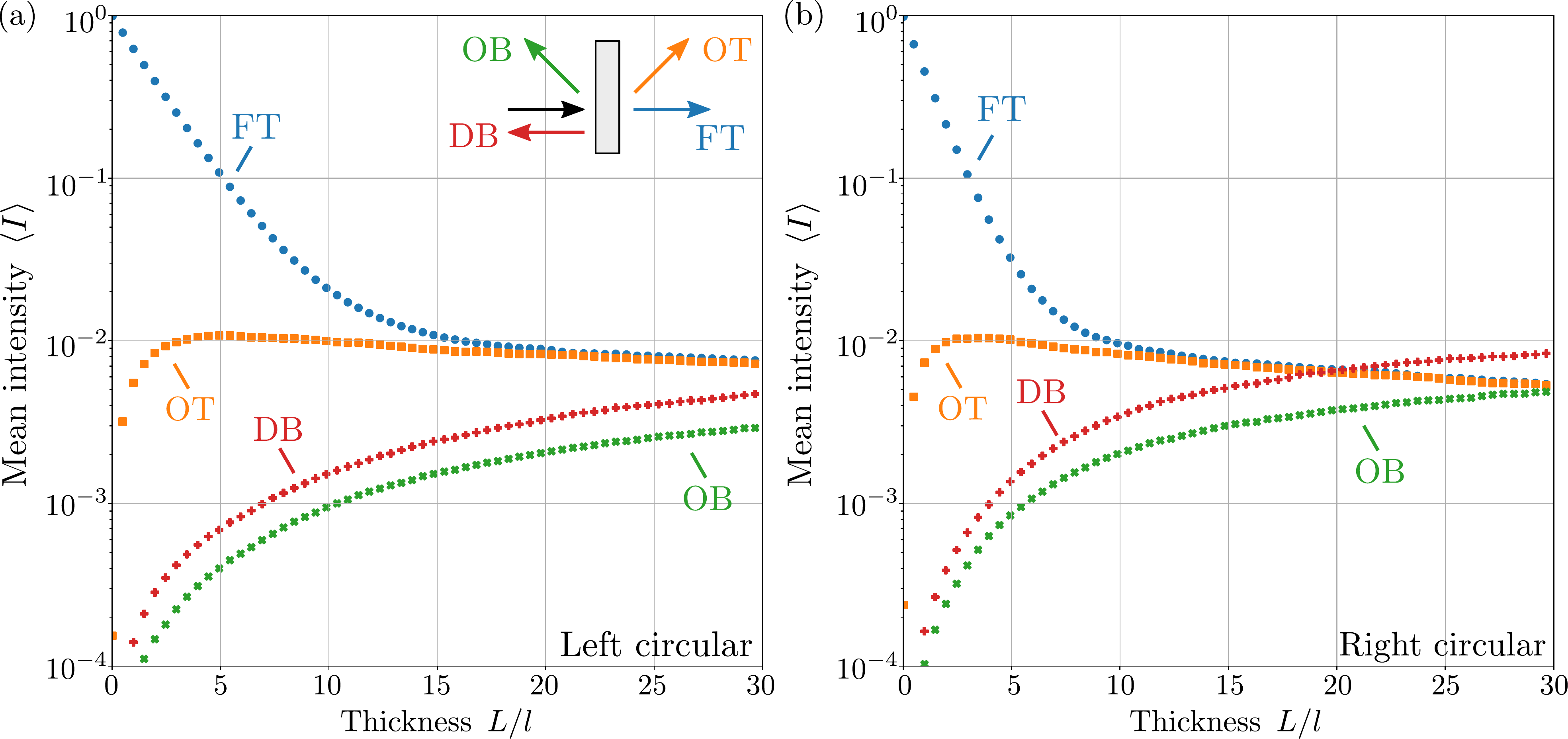}
	\caption{Mean intensity as a function of thickness for size parameter $x=4$ and incident (a) left and (b) right circular polarization. The intensity is shown in four different outgoing modes: forward transmission (FT), oblique transmission (OT), oblique backscattering (OB) and direct backscattering (DB). A visual aid is provided in (a).}\label{chiralint}
\end{figure}

Figures \ref{chiralint}(a) and \ref{chiralint}(b) show the mean scattered intensity for chiral spheres with size parameter $x=4$ for incident left handed circularly polarized light (LHC) and right handed circularly polarized light (RHC) respectively. While the overall trends closely resemble those in Figure \ref{mean_intensity_mie}, there is now a clear polarization dependence. As was the case with the isotropic spheres, much of the behaviour can be explained through consideration of scattering anisotropy. LHC, which is more preferentially forward scattered than RHC, decays slower in FT. For RHC, the mean intensity is correspondingly larger in the backscattering directions. Similar behaviour was seen for size parameter $x=1$.

\subsubsubsection{Degree of polarization}

\begin{figure}[t!]
	\centering\includegraphics[width =\textwidth]{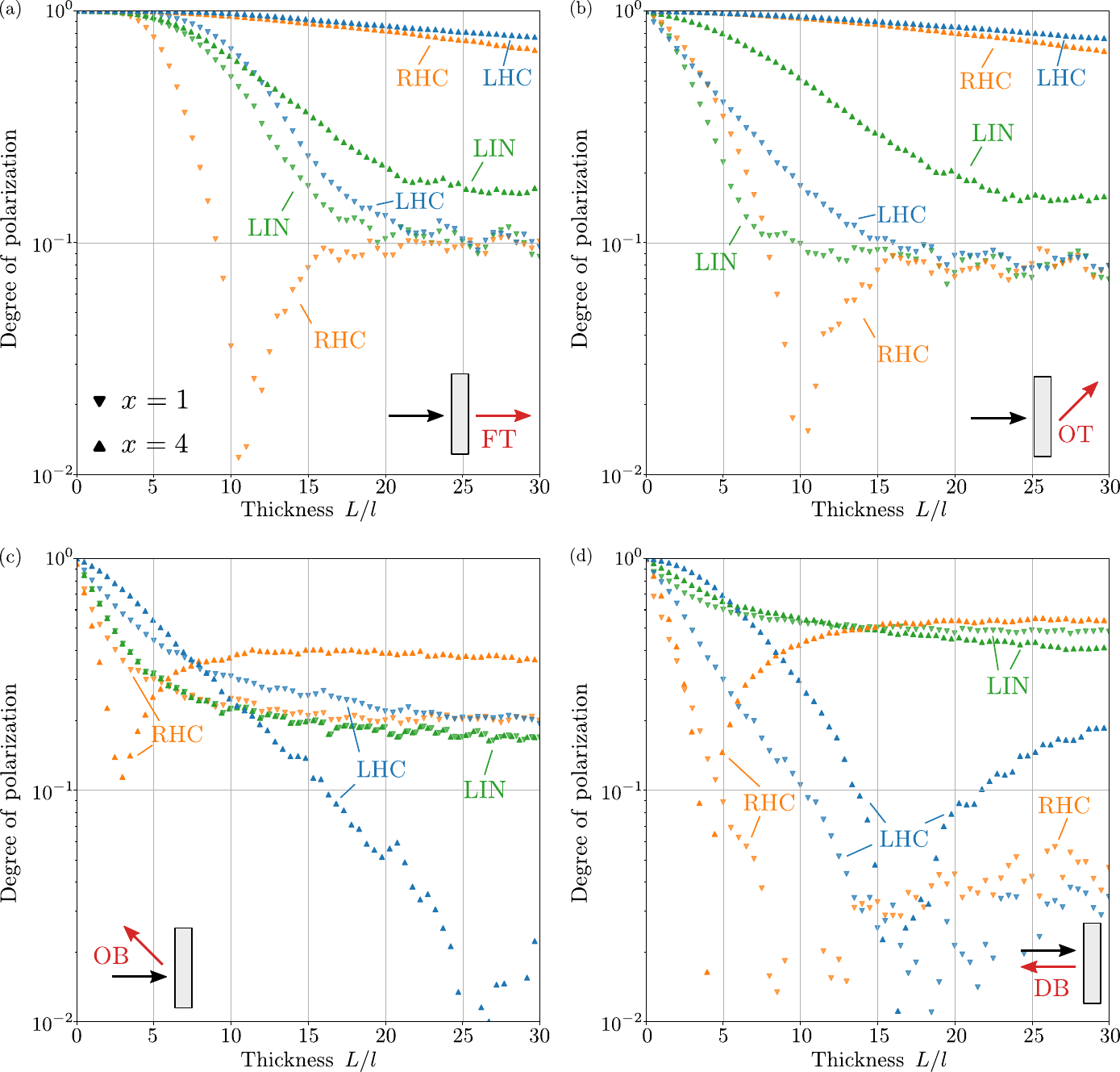}
	\caption{DoP as a function of thickness for incident linearly polarized light (LIN), left handed circularly polarized light (LHC) and right handed circularly polarized light (RHC) for size parameters $x=1$ ($\blacktriangledown$ markers) and 4 ($\blacktriangle$ markers) in outgoing modes (a) forward transmission (FT), (b) oblique transmission (OT), (c) oblique backscattering (OB) and (d) direct backscattering (DB).}\label{chiralgraphf}
\end{figure}

The DoP statistics for chiral spheres of size parameters $x=1$ and 4 (indicated by downward and upward pointing triangles respectively) are shown in Figure \ref{chiralgraphf}. We have included three different incident polarization states: LHC, RHC and LIN, the last of which refers to incident linearly polarized light. The trends we see are similar to those for isotropic spheres in Figure \ref{dopgraphs}, but with a few interesting differences. In Figure \ref{chiralgraphf}(a), a dip in the DoP can now be seen in FT for RHC and $x=1$. For isotropic spheres, these dips in the DoP were only present in reflection for larger spheres. We can explain this phenomenon however by invoking a similar argument to before. For RHC and a thin medium, the distribution of scattered polarization states on the Poincaré sphere was sharply peaked at the pole corresponding to RHC. For large thicknesses, however, we found that this distribution transitioned to one that was relatively isotropic, but with a slight concentration towards the LHC pole due to the particle chirality. Thus, as before, in transitioning between these two distributions, there exists an intermediate thickness at which the DoP attains a minimum value. This does not occur for incident LHC, as the initial distribution of scattered states is already concentrated at the LHC pole and no such transition occurs as thickness increases. For incident LIN, the distribution is initially focused at a point on the equator of the Poincaré sphere and, in transitioning towards a distribution focused at the LHC pole, there is no intermediate thickness at which the distribution is isotropic across the entire sphere. Therefore, no such dip in the DoP occurs. We note that we also expect a dip in the DoP to occur for incident RHC and $x=4$, but as the DoP decay rate is small for this size parameter, the medium is not thick enough, even at $30l$, for the dip to occur. In OT, as shown in Figure \ref{chiralgraphf}(b), we see that the behaviour resembles FT in the same way that Figure \ref{dopgraphs}(b) resembles Figure \ref{dopgraphs}(a).

In Figure \ref{chiralgraphf}(c), we see that for $x=1$ the DoP behaviour is similar to that of Figure \ref{dopgraphs}(c), but note that the DoP decays more quickly for RHC than for LHC. All three incident polarization states settle on similar limiting DoP values, with LHC and RHC $\sim 0.2$ and LIN $\sim 0.17$. For $x=4$, dips in the DoP are again visible for RHC and LHC. Unlike in Figure \ref{chiralgraphf}(a), these dips arise due to the flipping or preservation of helicity for different scattered field contributions, as was the case in Figure \ref{dopgraphs}(c). For LHC, which scatters more anisotropically, this dip occurs at a larger thickness ($L/l \sim 27$) than for RHC ($L/l \sim 3$). In Figure \ref{chiralgraphf}(d), we see that while linearly polarized light retains a large DoP for large thicknesses irrespective of particle size, dips in the DoP occur again for LHC and RHC and $x=4$. For $x=1$ and incident circularly polarized light, we also see dips in the DoP, but the exact trends are unclear. For DoP on the order of $10^{-2}$ a larger number of realizations than was used in this work is required for good numerical convergence.

\section{Conclusion}
To conclude, we have presented a method for randomly generating scattering matrices for sparse, complex media that incorporates the polarization properties of light, scattering anisotropy and the physical constraints of unitarity and reciprocity. Furthermore, we are able to model random media in the multiple scattering regime using a matrix cascade, only requiring knowledge of the single scattering properties of the particles contained within the medium. 

We have validated our model by reproducing known behaviour for systems consisting of randomly distributed spherical particles, such as the dependence of the rate of depolarization on the incident polarization state. We have also shown that some of the polarization statistics of our scattering matrices in the large thickness limit can be related to those of random Gaussian matrices and diagonal blocks of matrices drawn from the circular orthogonal ensemble. We have demonstrated the flexibility of our approach by considering the example of a medium containing chiral particles, for which we found that the polarization properties of the scattered field depend on the helicity of the incident polarization state. We were able to analyze the more intricate details of the rate of decay of DoP by considering the evolution of scattered polarization state distributions on the Poincaré sphere, which is easily done in our framework given that we have access to the entire scattering matrix. In addition to the data presented here, other possible studies include analyzing the polarization properties of the transmission eigenchannels and the polarization properties of correlations between different matrix blocks, such as, for example, the memory effect. We reserve these topics for future studies.

The biggest limitation of our model is the currently achievable angular resolution of the scattered field, as this directly influences the size of the scattering matrix, which, when large, requires a lot of memory and computation time when a large number of samples is required for the study of statistical quantities. Generation of individual scattering matrices, however, is very fast, taking only seconds or minutes, depending on the medium thickness and number of modes. We therefore envisage that our method will serve as a complement to the already existing Monte Carlo techniques and may prove advantageous in certain applications, particularly where correlations between different matrix elements are of interest.

\appendix
\section{Covariances and pseudo-covariances of scattering matrix elements}
\label{appendix:a}
Table~\ref{tab:appA} contains a list of expressions for the covariances and pseudo-covariances of the elements of the scattering matrix. Referring to the first column of Table~\ref{tab:appA}, type `Regular' refers to the covariance of the form $\langle \bar{B}_{(j,i)ba}\bar{B}^{*}_{(v,u)dc}\rangle - \langle \bar{B}_{(j,i)ba}\rangle \langle\bar{B}^{*}_{(v,u)dc}\rangle$, where $\bar{B}$ denotes an arbitrary block of the scattering matrix (i.e. one of $\bar{r}$,$\bar{t}$,$\bar{t}'$ or $\bar{r}'$). Type `Pseudo' refers to the pseudo-covariance of the form $\langle \bar{B}_{(j,i)ba}\bar{B}^{*}_{(v,u)dc}\rangle - \langle \bar{B}_{(j,i)ba}\rangle \langle\bar{B}^{*}_{(v,u)dc}\rangle$. All symbols are as defined in the main text.

\begin{table}[h!]\label{tableformulae}
		\caption{Summary of the regular and pseudo covariances of the elements of the scattering matrix.\label{tab:appA}}
	\begin{tabular}{@{}lll@{}}
		\toprule
		\multicolumn{1}{c}{Type}                                         & \multicolumn{1}{c}{Block $\bar{B}$}& \multicolumn{1}{c}{Expression}                                                                                                               \\ \cmidrule(lr){1-1}\cmidrule(lr){2-2}\cmidrule(lr){3-3}
		\multicolumn{1}{c}{Regular} & \multicolumn{1}{c}{$\bar{t}$}   & $\delta^R C_{ijuv}\langle A^{t}_{(j,i)ba}A^{t*}_{(j,i)dc}\rangle \sinc(\frac{L}{2}(k_{iz}- k_{jz} -k_{uz} + k_{vz}))$    \\
		\multicolumn{1}{c}{}                         & \multicolumn{1}{c}{$\bar{r}$}   & $\delta^R C_{ijuv}\langle A^{r}_{(j,i)ba}A^{r*}_{(j,i)dc}\rangle \sinc(\frac{L}{2}(k_{iz}+ k_{jz} -k_{uz} - k_{vz}))$    \\
		\multicolumn{1}{c}{}                         & \multicolumn{1}{c}{$\bar{t'}$}  & $\delta^R C_{ijuv}\langle A^{t'}_{(j,i)ba}A^{t'*}_{(j,i)dc}\rangle \sinc(\frac{L}{2}(-k_{iz}+ k_{jz} +k_{uz} - k_{vz}))$ \\
		\multicolumn{1}{c}{}                         & \multicolumn{1}{c}{$\bar{r'}$}  & $\delta^R C_{ijuv}\langle A^{r'}_{(j,i)ba}A^{r'*}_{(j,i)dc}\rangle \sinc(\frac{L}{2}(-k_{iz}- k_{jz} +k_{uz} - k_{vz}))$ \\
		\multicolumn{1}{c}{Pseudo}                      & \multicolumn{1}{c}{$\bar{t}$}   & $\delta^P C_{ijuv}\langle A^{t}_{(j,i)ba}A^{t}_{(j,i)dc}\rangle \sinc(\frac{L}{2}(k_{iz}- k_{jz} +k_{uz} - k_{vz}))$     \\
		& \multicolumn{1}{c}{$\bar{r}$}  & $\delta^P C_{ijuv}\langle A^{r}_{(j,i)ba}A^{r}_{(j,i)dc}\rangle \sinc(\frac{L}{2}(k_{iz}+ k_{jz} +k_{uz} + k_{vz}))$     \\
		& \multicolumn{1}{c}{$\bar{t'}$}  & $\delta^P C_{ijuv}\langle A^{t'}_{(j,i)ba}A^{t'}_{(j,i)dc}\rangle \sinc(\frac{L}{2}(-k_{iz}+ k_{jz} -k_{uz} + k_{vz}))$  \\
		& \multicolumn{1}{c}{$\bar{r'}$}  & $\delta^P C_{ijuv}\langle A^{r'}_{(j,i)ba}A^{r'}_{(j,i)dc}\rangle \sinc(\frac{L}{2}(-k_{iz}- k_{jz} -k_{uz} - k_{vz}))$  \\ \bottomrule
	\end{tabular}
\end{table}
\section{Composition law for scattering matrices}
\label{appendix:b}
Suppose two slabs, $\mathcal{S}_1$ and $\mathcal{S}_2$, with planar faces perpendicular to the $z$-axis are arranged such that $\mathcal{S}_1$ is to the left of $\mathcal{S}_2$, i.e. $z_1 < z_2$ where $z_1$ and $z_2$ are the $z$ coordinates of the centers of the slabs. If $\mathcal{S}_1$ and $\mathcal{S}_2$ have scattering matrices $\mathbf{S}_1$ and $\mathbf{S}_2$ where 
\begin{align}
\mathbf{S}_1 = \begin{pmatrix}
	\mathbf{r}_1 &\mathbf{t}'_1 \\
	\mathbf{t}_1 & \mathbf{r}'_1
\end{pmatrix} \quad\mathrm{and}\quad \mathbf{S}_2 = \begin{pmatrix}
\mathbf{r}_2 &\mathbf{t}'_2 \\
\mathbf{t}_2 & \mathbf{r}'_2
\end{pmatrix},
\end{align}
then the scattering matrix $\mathbf{S}$ for the overall system composed of $\mathcal{S}_1$ and $\mathcal{S}_2$ is given by 
\begin{align}
\mathbf{S} = \begin{pmatrix}
	\mathbf{r} &\mathbf{t}' \\
	\mathbf{t} & \mathbf{r}'
\end{pmatrix} = \begin{pmatrix}
\mathbf{r}_1 + \mathbf{t}'_1\mathbf{r}_2\mathbf{Q}\mathbf{t}_1 & (\mathbf{J}_{2N_k +1}\otimes \bm\sigma_z)(\mathbf{t}_2\mathbf{Q}\mathbf{t}_1)^\mathrm{T}(\mathbf{J}_{2N_k +1}\otimes \bm\sigma_z)\\
\mathbf{t}_2\mathbf{Q}\mathbf{t}_1 & \mathbf{r}'_2 + \mathbf{t}_2\mathbf{Q}\mathbf{r}'_1\mathbf{t}'_2 
\end{pmatrix},
\end{align}
where $\mathbf{Q} = (\mathbb{I}-\mathbf{r}'_1\mathbf{r}_2)^{-1}$, $\otimes$ is the Kronecker product, $\bm\sigma_z = \mathrm{diag}(1,-1)$ and $\mathbf{J}_n$ is the $n\times n$ exchange matrix containing 1s on its anti-diagonal and 0s elsewhere.

\section{Scattering and transfer matrices centred at arbitrary positions}
\label{appendix:propagator}
Suppose that a slab of thickness $\Delta L$ is centred at $L_0$ so that the $z$ coordinate of the position of any particular particle within the slab is confined to the interval $[L_0 - \Delta L/2, L_0 + \Delta L/2]$. Inspecting Eq. (\ref{t}), it can be seen that if such a slab has transmission matrix block $\bar{\mathbf{t}}^{L_0}_{(j,i)}$, then $\bar{\mathbf{t}}^{L_0}_{(j,i)} = \bar{\mathbf{t}}^{0}_{(j,i)}\exp[i(k_{iz} - k_{jz})L_0]$, where $\bar{\mathbf{t}}^{0}_{(j,i)}$ describes a medium identical to that described by $\bar{\mathbf{t}}^{L_0}_{(j,i)}$, but for which the $z$ coordinate of the position of each particle has been translated by $L_0$ so that each particle is now confined to the interval $[-\Delta L/2, \Delta L/2]$ centred at the origin. Consideration of the block structure of $\bar{\mathbf{t}}^{L_0}$, which is the full transmission matrix for scattering medium centred at $z=L_0$, then leads to the equation $\bar{\mathbf{t}}^{L_0} = \Lambda_{-}^{L_0}\bar{\mathbf{t}}^0\Lambda_{+}^{L_0}$, where
\begin{align}
	\Lambda_+^{L_0} = \begin{pmatrix}
		e^{ik_{-N_k z}L_0} & \hdots & 0 \\
		\vdots & \ddots & \vdots \\
		0 & \hdots & e^{ik_{N_k z}L_0} 
	\end{pmatrix}\otimes \begin{pmatrix}
	1 & 0 \\
	0 & 1
\end{pmatrix}\label{prop1}
\end{align}
and $\Lambda_-^{L_0} = (\Lambda_+^{L_0})^*$. On the right hand side of Eq. (\ref{prop1}), the arguments of the exponentials in the first matrix run through all modes in the set $K$ in order. Similar reasoning for the other blocks of the scattering matrix leads to Eq. (\ref{propagatorEq}) in the main text, where $\Lambda_{\pm}^{L_0} = \mathrm{diag}(\Lambda^{L_0}_+, \Lambda^{L_0}_-)$ and $\Lambda^{L_0}_{\mp} = (\Lambda^{L_0}_{\pm})^*$.

Suppose now that a series of $N$ scattering layers are situated with centres located at (from left to right) $L_1, L_2, \cdots L_N$ and let $\bar{\mathbf{M}}^{L_i}_{i}$ denote the transfer matrix for the $i$'th layer. Using Eq. (\ref{propagatorEq}), the overall transfer matrix is given by
\begin{align}
	\begin{split}
	\bar{\mathbf{M}} &= \bar{\mathbf{M}}^{L_N}_{N}\hdots\bar{\mathbf{M}}^{L_3}_{3}\bar{\mathbf{M}}^{L_2}_{2}\bar{\mathbf{M}}^{L_1}_{1}\\
	&= \Lambda_{\mp}^{L_N}\bar{\mathbf{M}}^{0}_{N}\Lambda_{\pm}^{L_N}\hdots\Lambda_{\mp}^{L_3}\bar{\mathbf{M}}^{0}_{3}\Lambda_{\pm}^{L_3}\Lambda_{\mp}^{L_2}\bar{\mathbf{M}}^{0}_{2}\Lambda_{\pm}^{L_2}\Lambda_{\mp}^{L_1}\bar{\mathbf{M}}^{0}_{1}\Lambda_{\pm}^{L_1}\\
	& = \Lambda_{\mp}^{L_N}\bar{\mathbf{M}}^{0}_{N}\hdots\bar{\mathbf{M}}^{0}_{3}\Lambda_{\pm}^{L_3-L_2}\bar{\mathbf{M}}^{0}_{2}\Lambda_{\pm}^{L_2-L_1}\bar{\mathbf{M}}^{0}_{1}\Lambda_{\pm}^{L_1}
	\end{split},\label{propagator2}
\end{align}
where, as always, a superscript $0$ denotes the corresponding transfer matrix when the slab is centred at the origin. Deriving the final line of Eq. (\ref{propagator2}) makes use of the identity $\Lambda_{\pm}^{L_2}\Lambda_{\mp}^{L_1} = \Lambda_{\pm}^{L_2-L_1}$, which follows trivially from the definitions. In the special case $L_1 = 0$ and $L_{i+1} - L_{i} = \Delta L$ for $1 \leq i \leq N-1$, as would be the case for contiguous slabs of equal thicknesses $\Delta L$, Eq. (\ref{propagator2}) can be written in the form
\begin{align}
	\bar{\mathbf{M}} = \Lambda^{N\Delta L}_{\mp} \prod_{i=1}^{N}\Lambda_{\pm}^{\Delta L}\bar{\mathbf{M}}_i^0.\label{propfinal}
\end{align}
Therefore, a transfer matrix for a medium of thickness $N \Delta L$ can be computed by cascading $N$ matrices of the form $\Lambda_{\pm}^{\Delta L}\bar{\mathbf{M}}^0$, where $\bar{\mathbf{M}}^0$ can be randomly generated as discussed in the main text. Note that the final matrix $\Lambda^{N\Delta L}_{\mp}$ outside of the product in Eq. (\ref{propfinal}) imparts global phase terms onto each $2\times 2$ block of $\bar{\mathbf{M}}$ (and $\bar{\mathbf{S}}$) and therefore does not alter any of the intensity or polarization statistics of the random matrix given by the product. 

\section{Computation of single particle scattering matrices}
\label{appendix:rotation}
Consider a particular pair of incident and outgoing plane waves with wavevectors $\mathbf{k}_i$ and $\mathbf{k}_j$ respectively. Let $\mathbf{e}_{ki}$, $\mathbf{e}_{\phi i}$, $\mathbf{e}_{\theta i}$ $\mathbf{e}_{kj}$, $\mathbf{e}_{\phi j}$ and $\mathbf{e}_{\theta j}$ be the associated spherical polar vectors as defined as in Eq. (\ref{polar}). The vectors $\mathbf{e}_{ki}$ and $\mathbf{e}_{kj}$ define the scattering plane, whose unit normal vector is given by $\mathbf{e}_{\perp} =(\mathbf{e}_{ki} \times \mathbf{e}_{kj})/|\mathbf{e}_{ki} \times \mathbf{e}_{kj}|$. We then define the vectors $\mathbf{e}_{\parallel i} = \mathbf{e}_{\perp} \times \mathbf{e}_{ki}$ and $\mathbf{e}_{\parallel j} = \mathbf{e}_{\perp} \times \mathbf{e}_{kj}$ so that ($\mathbf{e}_{\parallel i}, \mathbf{e}_{\perp i}, \mathbf{e}_{ki}$) and ($\mathbf{e}_{\parallel j}, \mathbf{e}_{\perp j}, \mathbf{e}_{kj}$) form right-handed triads. In the case that $\mathbf{e}_{ki}$ and $\mathbf{e}_{kj}$ are parallel, we take $\mathbf{e}_{\parallel i } = \mathbf{e}_{\theta i}$, $\mathbf{e}_{\parallel j } = \mathbf{e}_{\theta j}$ and $\mathbf{e}_{\perp i} = \mathbf{e}_{\perp j} = \mathbf{e}_{\phi i}$. 

\begin{figure}[t!]
	\centering\includegraphics[width =0.3\textwidth]{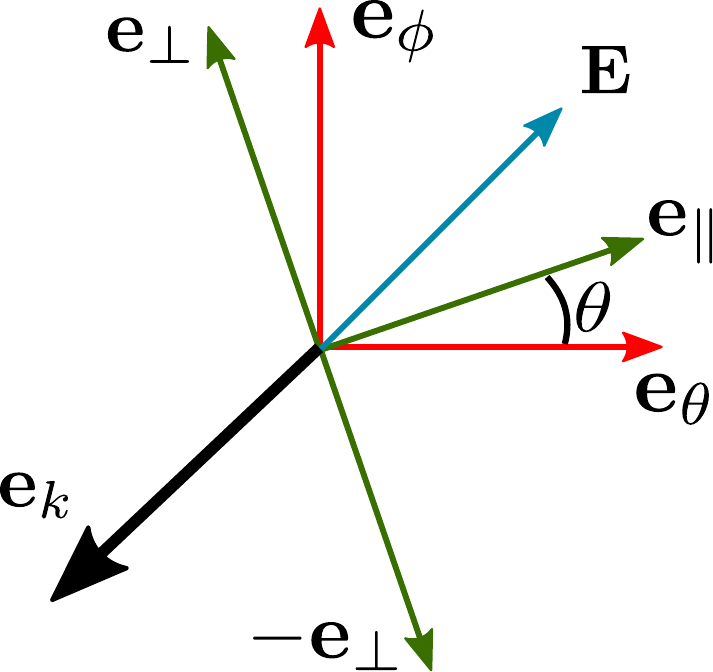}
	\caption{Vectors used in scattering calculations. The vectors  $\mathbf{E}, \mathbf{e}_\theta, \mathbf{e}_\phi, \mathbf{e}_\parallel$ and  $\mathbf{e}_\perp$ all lie in the plane perpendicular to $\mathbf{e}_k$. The angle $\theta$ is positive in the diagram.}\label{angles}
\end{figure}

Consider now the incident wavevector $\mathbf{k}_i$ and let us temporarily drop the subscript $i$. the vectors $\mathbf{e}_\theta, \mathbf{e}_\phi, \mathbf{e}_\parallel$ and $\mathbf{e}_\perp$ all lie in the same plane with unit normal vector given by $\mathbf{e}_k$. In general, however, the vectors $\mathbf{e}_\theta$ and $\mathbf{e}_\phi$ will not align with $\mathbf{e}_\parallel$  and $\mathbf{e}_\perp$. Let $\theta$ be the angle between $\mathbf{e}_\theta$ and $\mathbf{e}_\parallel$, defined such that $-\pi < \theta < \pi$, where $\theta > 0$ if $(\mathbf{e}_\theta \times \mathbf{e}_\parallel)/|\mathbf{e}_\theta \times \mathbf{e}_\parallel| = \mathbf{e}_k$  (i.e. $\mathbf{e}_\parallel$ is an anti-clockwise rotation of $\mathbf{e}_\theta$ about $\mathbf{e}_k$) and $\theta < 0 $ if $(\mathbf{e}_\theta \times \mathbf{e}_\parallel)/|\mathbf{e}_\theta \times \mathbf{e}_\parallel| = -\mathbf{e}_k$ (i.e. $\mathbf{e}_\parallel$ is a clockwise rotation of $\mathbf{e}_\theta$ about $\mathbf{e}_k$). See Figure \ref{angles} for a graphical representation of these vectors, along with the electric field vector $\mathbf{E}$, which also lies in the same plane.

Given $\theta$, the electric field vector, which can be written as $\mathbf{E} = (E_\theta, E_\phi)^\mathrm{T}$ with respect to the basis vectors $\mathbf{e}_\theta$ and $\mathbf{e}_\phi$, can be transformed to $\mathbf{E} = (E_\parallel, E_\perp)^\mathrm{T}$ with respect to $\mathbf{e}_\parallel$ and $\mathbf{e}_\perp$ by
\begin{align}
\begin{pmatrix}
E_\parallel \\
E_\perp 
\end{pmatrix} = 
\mathbf{R}(\theta)\begin{pmatrix}
E_\theta \\
E_\phi
\end{pmatrix},\quad\quad \mathbf{R}(\theta) = \begin{pmatrix}
\cos(\theta) &\sin(\theta) \\
-\sin(\theta) & \cos(\theta) 
\end{pmatrix}.
\end{align}
Note that conventions for the directions of the unit vectors described here are not consistent throughout the literature. For example, in Ref \cite{bohren2008absorption}, the normal to the scattering plane is taken to be $\mathbf{e}'_\perp = -\mathbf{e}_\perp$. In this case, the electric field component perpendicular to the scattering plane is given by $E'_\perp = -E_\perp$. Following the convention used in Ref \cite{bohren2008absorption}, it can ultimately be shown that
\begin{align}
\begin{pmatrix}
E_{\theta j} \\
E_{\phi j}
\end{pmatrix} = \mathbf{R}(-\theta_j)\bm\sigma_z\begin{pmatrix}
S_2 & S_3 \\
S_4 & S_1 
\end{pmatrix}\bm\sigma_z\mathbf{R}(\theta_i)\begin{pmatrix}
E_{\theta i} \\
E_{\phi i}
\end{pmatrix},\label{appendixCfinal}
\end{align}
where $S_1$, $S_2$, $S_3$ and $S_4$ are scattering coefficients defined with respect to the scattering plane and $\theta_i$ and $\theta_j$ are the angles between $\mathbf{e}_{\theta i}$, $\mathbf{e}_{\parallel i}$ and $\mathbf{e}_{\theta j}$, $\mathbf{e}_{\parallel j}$ respectively, following the sign convention as discussed. Finally, the matrix $\mathbf{A}_{(j,i)}^{t/r}$ is given by the product of the five matrices in Eq. (\ref{appendixCfinal}).

\section*{Acknowledgements}
This work was funded by the Royal Society (grant numbers RGF\textbackslash R1\textbackslash180052, UF150335 and URF\textbackslash R\textbackslash211029).

\section*{Disclosure statement}
The authors report there are no competing interests to declare.

\end{document}